\documentclass[fleqn,usenatbib]{mnras}

\usepackage[T1]{fontenc}
\usepackage{ae,aecompl}

\usepackage{graphicx}	
\usepackage{amsmath}	
\usepackage{amssymb}	

\usepackage{newtxtext,newtxmath}

\usepackage[usenames,dvipsnames]{xcolor}
\usepackage{physics}
\usepackage{tikz}
\usetikzlibrary{arrows,babel}
\usepackage[labelfont={bf},labelsep=endash]{caption}
\usetikzlibrary{positioning,arrows}
\usetikzlibrary{decorations.pathmorphing}
\usetikzlibrary{decorations.markings}
\tikzset{
momentum/.style={postaction={decorate},decoration={markings,mark=at position 1 with {\arrow{>}}}},
particle/.style={dashed
    },
photon/.style={decorate, 
    decoration={snake}},
    math/.style={draw, execute at begin node={$\displaystyle}, execute at end node={$}}
 }

\usepackage{pifont}  
\usepackage{enumitem}
\usepackage[normalem]{ulem}

\newcommand{\sub}[1]{_{\mathrm{#1}}}
\newcommand{\msun}{\mathrm{M\sub{\sun}}}
\newcommand{\nbody}{$N$-body }

\title[Shedding light on low-mass subhalo survival and annihilation luminosity]{Shedding light on low-mass subhalo survival and annihilation luminosity with numerical simulations}

\author[A. Aguirre-Santaella et al.]{
Alejandra Aguirre-Santaella,$^{1, 2}$\thanks{e-mail: alejandra.aguirre@uam.es}
Miguel A. S\'anchez-Conde,$^{1, 2}$ 
Go Ogiya,$^{3, 4, 5}$ 
\newauthor Jens St\"ucker,$^{6}$ 
Raul E. Angulo$^{6, 7}$
\\
$^{1}$ Instituto de F\'isica Te\'orica UAM-CSIC, Universidad Aut\'onoma de Madrid, C/ Nicol\'as Cabrera, 13-15, 28049 Madrid, Spain\\
$^{2}$ Departamento de F\'isica Te\'orica, M-15, Universidad Aut\'onoma de Madrid, E-28049 Madrid, Spain\\
$^{3}$ Institute for Astronomy, School of Physics, Zhejiang University, Hangzhou 310027, China \\
$^{4}$ Waterloo Centre for Astrophysics, University of Waterloo, Waterloo, ON N2L 3G1, Canada \\
$^{5}$ Department of Physics and Astronomy, University of Waterloo, 200 University Avenue West, Waterloo, Ontario N2L 3G1, Canada \\
$^{6}$ Donostia International Physics Center (DIPC), Manuel Lardizabal Ibilbidea, 4, 20018 Donostia, Gipuzkoa, Spain\\
$^{7}$ IKERBASQUE, Basque Foundation for Science, 48013, Bilbao, Spain\\}

\date{Accepted XXX. Received YYY; in original form ZZZ}

\pubyear{2021}

\begin{document}
\label{firstpage}
\pagerange{\pageref{firstpage}--\pageref{lastpage}}
\maketitle

\begin{abstract}
In this work, we carry out a suite of specially-designed numerical simulations to shed further light on dark matter (DM) subhalo survival at mass scales relevant for gamma-ray DM searches, a topic subject to intense debate nowadays. Specifically, we have developed and employed an improved version of DASH, a GPU \nbody code, to study the evolution of low-mass subhaloes inside a Milky Way-like halo with unprecedented accuracy, reaching solar-mass and sub-parsec resolution in our simulations. We simulate subhaloes with varying mass, concentration, and orbital properties, and consider the effect of the gravitational potential of the Milky Way galaxy itself. 
More specifically, we analyze the evolution of both the bound mass fraction and annihilation luminosity of subhaloes, finding that most subhaloes
survive until present time, 
even though in some cases they lose more than 99\% of their mass at accretion. Baryons in the host induce a much more severe mass loss, especially when the subhalo orbit is more parallel to the galactic disk. 
Many of these subhaloes cross the solar galactocentric radius, thus making it easier to detect their annihilation fluxes from  Earth. 
We find subhaloes orbiting a DM-only halo with a pericentre in the solar vicinity to lose 70-90\% of their initial annihilation luminosity at redshift zero, which increases up to 99\% when baryons are also included in the host. 
We find a strong relation between subhalo's mass loss and the effective tidal field at pericentre. Indeed, much of the dependence on concentration, orbital parameters, host potential and baryonic components can be explained through this single parameter. 
In addition to shedding light on the survival of low-mass galactic subhaloes,  our results can provide detailed predictions that will aid current and future quests for the nature of DM.
\end{abstract}





\section{Introduction}

There is strong evidence to believe that there should exist something else apart from the matter we are able to observe in the Universe. 
Indeed, there are completely independent cosmological and astrophysical observations that point that, if our theory of gravity is correct, the mass of the matter we 
can detect electromagnetically is not enough to explain certain phenomena, whilst adding a new matter component, dark matter (DM), they are possible~\citep{Bertone:2004pz, Garrett:2010hd, 2018RvMP...90d5002B, 2020A&A...641A...6P}.

Despite our efforts, the nature of the DM is yet unknown. There are three main different yet complementary methods to look for DM: direct production (using collider experiments in particle accelerators), direct detection (that look for traces of interactions between DM and baryonic matter at the laboratory)~\citep{2017IJMPA..3230006K} and indirect detection~\citep{2005MPLA...20.1021B}. The indirect detection method aims to observe the radiation (gamma-rays and neutrinos) and antimatter (e.g. positrons) produced by DM annihilation or decay into Standard Model particles which could be detected through spatial or terrestrial observatories
, such as H.E.S.S.~\citep{2004NewAR..48..331H}, MAGIC~\citep{2004ASPC..327...52F}, VERITAS~\citep{Weekes:2001pd}, \textit{Fermi}-LAT~\citep{fermiglast}, IceCube~\citep{Achterberg:2006md}, AMS~\citep{2008NIMPA.588..227B} 
and PAMELA~\citep{PICOZZA2007296}. 
A detection of these annihilation products might give a hint about DM properties~\citep{2011ARA&A..49..155P}.  
Moreover, all evidence we have on DM is astrophysical as of today, thus indirect searches are the only ones that have the potential not only to make the necessary connection between the nature of the DM and the astrophysical observations, but also to provide direct information about the actual DM distribution in the Universe.

Standard $\Lambda$CDM cosmology predicts a hierarchical procedure for structure formation, starting with low-mass virialized objects, or {\it haloes}, which later in time merge forming larger structures~\citep{2006Natur.440.1137S, 2012AnP...524..507F, 2019Galax...7...81Z}.
As a consequence, there is a huge amount of low-mass subhaloes inside larger haloes like our galaxy, the Milky Way (MW). The dwarf satellite galaxies are hosted by the most massive subhaloes, while there are also dark satellites (less massive subhaloes with no stars or gas at all) which do not possess visible counterparts. 

Using cosmological \nbody simulations with a large number of particles per virialized object and a high time and force resolution makes it possible to study the formation of cold DM haloes and their substructure in the non-linear regime in great detail~\citep{jurg1, 2020NatRP...2...42V, 2022LRCA....8....1A}.
Some of them 
are done assuming that all the matter is dark, that is, baryons are not included. Hence, they are collisionless \nbody simulations, and even though they are not so precise near the centre of the galaxies, they give an accurate solution of the idealized problem and are by far the best tool we have to understand structure formation and halo structural properties at present. Hydrodynamical simulations are also available nowadays~\citep{2014MNRAS.444.1518V, 2016MNRAS.457..844F, 2016MNRAS.457.1931S}, which include baryonic material inside the host, thus being more realistic. Nonetheless, basic properties of subhaloes such as their abundance, distribution and structure remain unclear for the less massive subhaloes due to the limited resolution in the simulations~\citep{2014MNRAS.442.3256A}. These simulations typically output subhaloes of at least one million solar masses~\citep{vlii_paper, 2008MNRAS.391.1685S, 2021MNRAS.506.4210I}, i.e. twelve orders of magnitude larger than the minimum halo mass expected in many DM scenarios. Also, finite numerical resolution implies that at least some subhaloes will be artificially destroyed in simulations.

Indeed, it is unclear whether small subhaloes will survive the strong tidal forces within their hosts since their accretion times to present~\citep{2003ApJ...584..541H, vandenBosch2018_analytic, vandenBosch2018_num_criteria}. Some authors claim that almost all subhalo disruption is of numerical origin and a bound remnant should always survive~\citep{vandenBosch2018_analytic, Ogiya2019, 2020MNRAS.491.4591E, 2021MNRAS.tmp.2848G, 2021arXiv211101148A, 2022arXiv220700604S}, while other studies suggest that the abundance of small subhaloes is severely reduced due to the effect of tidal forces and of other dynamical agents such as the presence of baryonic material~\citep{2017MNRAS.471.1709G, Kelley2019, 2021MNRAS.501.3558G, 2021MNRAS.507.4953G}. 
 There is about five times more DM than baryonic matter, hence the first one often governs the dynamics. Baryons are particularly important in the centres of large haloes, where galaxies form.

Both subhaloes hosting dwarf satellite galaxies and dark satellites are known to be excellent targets for gamma-ray DM searches since some of them may be close enough to yield large DM annihilation fluxes at Earth~\citep{2015PhRvL.115w1301A, Coronado_Blazquez2019, Coronado-Blazquez2019_2}. Also, the DM-annihilation flux is related to the annihilation luminosity, which is proportional to the DM density squared. Thus, the clumpy distribution of subhaloes will considerably boost the total DM annihilation in their host haloes, reaching values of up to a factor $\sim60$ for galaxy clusters~\citep{mascprada14, Moline17, Ando:2019xlm}. 
Note that having more resilient subhaloes would impact not only this boost computation but also almost every DM constraint obtained to date, as subhaloes are expected to  play a key role in almost every DM target.

Here, we carry out a suite of specially-designed numerical simulations to shed further light on subhalo survival at all mass scales relevant for DM searches. Specifically, we have employed the DASH\footnote{While DASH is actually the name of the simulation library, we are calling the code used in our work this way for simplicity.} simulation code~\citep{Ogiya2019} to study the evolution of subhaloes inside a MW-like halo with unprecedented accuracy. DASH is a fast tree-code 
optimised for GPU clusters which features both high performance and scalability. It simulates the dynamical evolution of subhaloes with the \nbody method and analytically describes the gravitational potential of the host. In this way, computational resources are focused on a single subhalo, which allows its simulation with extremely high force and mass resolution, which would not be possible in standard cosmological simulations.
More precisely, we will throw a subhalo inside the host and follow its dynamics under different initial configurations such as concentrations, masses, orbital parameters and accretion redshifts. We will also analyze the effect of taking into account the baryonic disk in the host potential.

Our work is expected to be particularly relevant for DM searches which, indeed, represent one of our ultimate goals. On one hand, we may get significantly larger DM fluxes at Earth from astrophysical objects, such as entire galaxies or galaxy clusters, if we can prove that a significant amount of small subhaloes survive the tidal forces they undergo since their accretion times till present time. This would also impact the computation of the subhalo boost, which could be now calculated in a more realistic way considering the actual abundance and properties of low-mass subhaloes. On the other, some of the surviving, tiny subhaloes closest to Earth would be excellent DM targets by themselves. In this sense, our suite of simulations and obtained results help enlightening the current debate on whether a considerable amount of subhaloes disrupt due to the tidal forces they experience or, on the contrary, we can still hope to look for them with our telescopes. 

The work is organized as follows. In Section~\ref{sec:sim_model}, we describe the code we have used and the modifications we have implemented for this work. The results of our study are depicted in Section~\ref{sec:resul}, giving special attention to two quantities, the bound mass fraction and the annihilation luminosity, both for runs without and with baryons. In Section~\ref{sec:discu}, we discuss our main findings and compare them to the results in the companion paper, \citet{2022arXiv220700604S}, where we present an analytical model that treats tidal stripping in the adiabatic limit to predict lower bounds on the asymptotic remnants of subhaloes. Finally, we conclude in Section~\ref{sec:conclu}.

\section{Simulation Model} \label{sec:sim_model}
We simulate the dynamical evolution of a DM subhalo orbiting within the MW potential, which consists of a DM host halo, stellar and gas disks, and a bulge. The subhalo is modelled as an \nbody system, while a time-evolving analytical potential is employed to model the MW. In this Section, we describe our simulation model and parameter choice.

\subsection{Subhalo} 
\label{ssec:subhalo}

In this study, we consider subhaloes that do not host any stars, and thus they purely consist of DM. Due to the cosmic UV background radiation, star formation in haloes with a virial mass $\,
\lesssim 10^8~\msun$ is suppressed and the gas within such haloes evaporates \citep[e.g.][]{Bullock2000,Okamoto2008}. While we employ the subhalo mass of $m\sub{sub} = 10^6~\msun$ in our main simulations, the simulation results can be, in principle, scaled down to arbitrarily small halo masses~\citep{2022arXiv220700604S}. Specifically, in this work we have tested subhalo masses down to $1~\msun$ (see Section \ref{sec:fbbary}).

We suppose that prior to accretion, the subhalo is spherical and follows the Navarro-Frenk-While (NFW) density profile~\citep{Navarro1997}, 
\begin{equation}
    \rho(r) = 4\rho\sub{s} (r/r\sub{s})^{-1} (1+r/r\sub{s})^{-2},
        \label{eq:nfw}
\end{equation}
where $r$ represents the distance from the centre of the halo, and $\rho\sub{s}$ 
and $r\sub{s}$ are the scale density and radius, respectively. The pair of the structural parameters ($\rho\sub{s}$ and $r\sub{s}$) can be derived from another pair of parameters, and we employ a pair of the virial mass\footnote{This includes only DM mass, as well as (1) in  Table~\ref{tab:mwpot}, so that the total host mass is smaller in runs without baryons.}, $M\sub{200}$, and the halo concentration, $c$, to specify the internal structure of the DM halo in what follows. The virial mass of the halo is given as
\begin{equation}
    M\sub{200} \equiv (800 \pi / 3) \rho\sub{crit}(z) r\sub{200}^3,
        \label{eq:m200}
\end{equation}
where $\rho\sub{crit}(z)$ is the critical density of the Universe at redshift $z$, and $r\sub{200}$ is the virial radius of the halo within which the mean density corresponds to $200 \rho\sub{crit}(z)$. The halo concentration is defined as $c \equiv r\sub{200}/r\sub{s}$. 

The initial positions of \nbody particles with respect to the centre of the subhalo are stochastically drawn by using the acceptance-rejection sampling method~\citep{Press2002}. We draw $r$ of a particle based on \autoref{eq:nfw} 
and its 3D position vector is specified with a randomly drawn unit vector. We stochastically draw the particle energy, $e$, based on the phase-space distribution function, $f(e)$. Here, $f(e)$ is numerically computed using the~\cite{Eddington1916} formula. Then we compute the velocity of the particle, $v$, with $e$ and the gravitational potential of the subhalo. The subhalo has an isotropic velocity dispersion
\footnote{While the velocity structure in the halo outskirts is radially biased, that in the halo centre ($r 
\lesssim r\sub{s}$) is almost isotropic \citep[e.g.][]{2009MNRAS.399..812W,2010MNRAS.402...21N}. 
 The radially biased velocity structure could enhance the mass loss rate at the beginning of simulations, when the subhalo outskirt is tidally stripped. However, our model should be fine in the later phase in which the subhalo centre has suffered from the tidal effects.} since $f(e)$ is assumed to depend only on the energy, and we specify the 3D velocity vector of the particle with another randomly drawn unit vector.

\subsection{The host potential} 
\label{ssec:host_pot}

The host potential is composed of a spherical DM host halo and the MW galaxy that consists of stellar and gas disks and a spherical bulge. The structural parameters of each component evolve with time, based on the empirical relations from cosmological simulations and observations. We input the masses of the DM halo and baryons and parameters introducing their spatial scales at $z=0$. 
These are all summarized in \autoref{tab:mwpot}. The centre of the host potential is fixed at the origin of the coordinate system in the entire simulations, i.e. simulations are performed in the host-centric frame. Note that dynamical friction is absent in our simulations as the host is modelled with an analytical potential. Neglecting dynamical friction is justified for the low mass subhaloes we explore in this paper, as the deceleration of dynamical friction is proportional to the subhalo mass  \citep{Chandrasekhar1943}.

\begin{table}
	\centering
	\caption{Input parameters for the host potential. They are the values at $z=0$. Description for each row: 
	(1) mass of the DM host halo, 
	(2) mass of the stellar disk, 
	(3) scale radius of the stellar disk, 
	(4) scale height of the stellar disk,  
	(5) mass of the gas disk, 
	(6) scale radius of the gas disk, 
	(7) scale height of the gas disk,
	(8) mass of the bulge, 
	(9) scale length of the bulge. 
	Note that the spatial scale $r_\mathrm{s}$ 
	of the DM host halo 
	is determined with Equation~\autoref{eq:m200} and the concentration-mass-redshift relation. 
	Parameters taken from~\citet{Kelley2019}.
	}
	\label{tab:mwpot}
	\begin{tabular}{ccc} 
		\hline
		 (1) & $M\sub{200,host}$  & $1.0 \cdot 10^{12}$ [$\msun$] \\
		 (2) & $M\sub{d,stellar}$ & $4.1 \cdot 10^{10}$ [$\msun$] \\
		 (3) & $R\sub{d,stellar}$ & $2.5$ [kpc]     \\
		 (4) & $h\sub{d,stellar}$ & $0.35$ [kpc]     \\
		 (5) & $M\sub{d,gas}$     & $1.9 \cdot 10^{10}$ [$\msun$] \\
		 (6) & $R\sub{d,gas}$     & $7.0$ [kpc]     \\
		 (7) & $h\sub{d,gas}$     & $0.08 $ [kpc]     \\
		 (8) & $M\sub{bulge}$     & $9.0 \cdot 10^{9}$ [$\msun$] \\
		 (9) & $R\sub{bulge}$     & $0.5 $ [kpc]     \\
		\hline
	\end{tabular}
\end{table}

\subsubsection{Host halo} 
\label{sssec:host_halo}

The DM host halo is assumed to be spherical in the entire simulation and is modelled with an analytical NFW potential. The virial mass of the host halo potential grows with the model for the mass assembly history of DM haloes by~\cite{Correa2015}. The concentration of the host halo is derived with the concentration-mass-redshift relation by~\cite{2016MNRAS.460.1214L}. 
These structural parameters are updated at every timestep in the simulation~\citep[explained in Section~\ref{ssec:num_tech}; see also][]{Ogiya2021}. Note that the DASH simulations performed by~\cite{Ogiya2019} employed a static NFW potential to model the DM host halo.

\subsubsection{The MW potential} 
\label{sssec:mw}

As an important update from the original DASH simulations~\citep{Ogiya2019}, our simulations now can include not only the DM host halo but also the baryonic components, the central bulge and stellar and gas disks, of the host potential following the recipe of~\cite{Kelley2019}. We employ a~\cite{Hernquist1990} potential to represent the central bulge,
\begin{equation}\label{eq-pot-hernq}
    \Phi_\mathrm{Hq}(r) = -\frac{G\,M\sub{bulge}}{r+R\sub{bulge}}\,,
\end{equation}
where $M\sub{bulge}$ and $R\sub{bulge}$ are the mass and the scale radius of the bulge, respectively. Each of the stellar and gas disks is supposed to be an exponential disk, and the mass, scale radius and scale height of the exponential disks at the present time are listed in \autoref{tab:mwpot}. \cite{Flynn1996} showed that an exponential disk is well approximated by combining separated Miyamoto-Nagai (MN) disks~\citep{Miyamoto1975} whose potential is given as
\begin{equation}\label{eq-pot-mn}
    \Phi_\mathrm{MN}(r,z) = -\frac{G\,M\sub{MN}}{\sqrt{r^2+(\sqrt{z^2+b^2}+a)^2}},
\end{equation}
where $M\sub{MN}$ is the mass of the MN disk and $a$ and $b$ are the MN disk scale radius and thickness, respectively. Following the prescription by~\cite{Smith2015}, an exponential disk is approximated with three separated MN disks, and the exponential disk parameters ($M\sub{d}$, $R\sub{d}$ and $h\sub{d}$) are converted to the parameters of three MN disks (three sets of $M\sub{MN}$, $a$ and $b$). Since two exponential disks (stellar and gas) are included in the simulations, we have six MN disks in total.

\begin{figure}
\begin{center}
\includegraphics[width=.94\columnwidth]{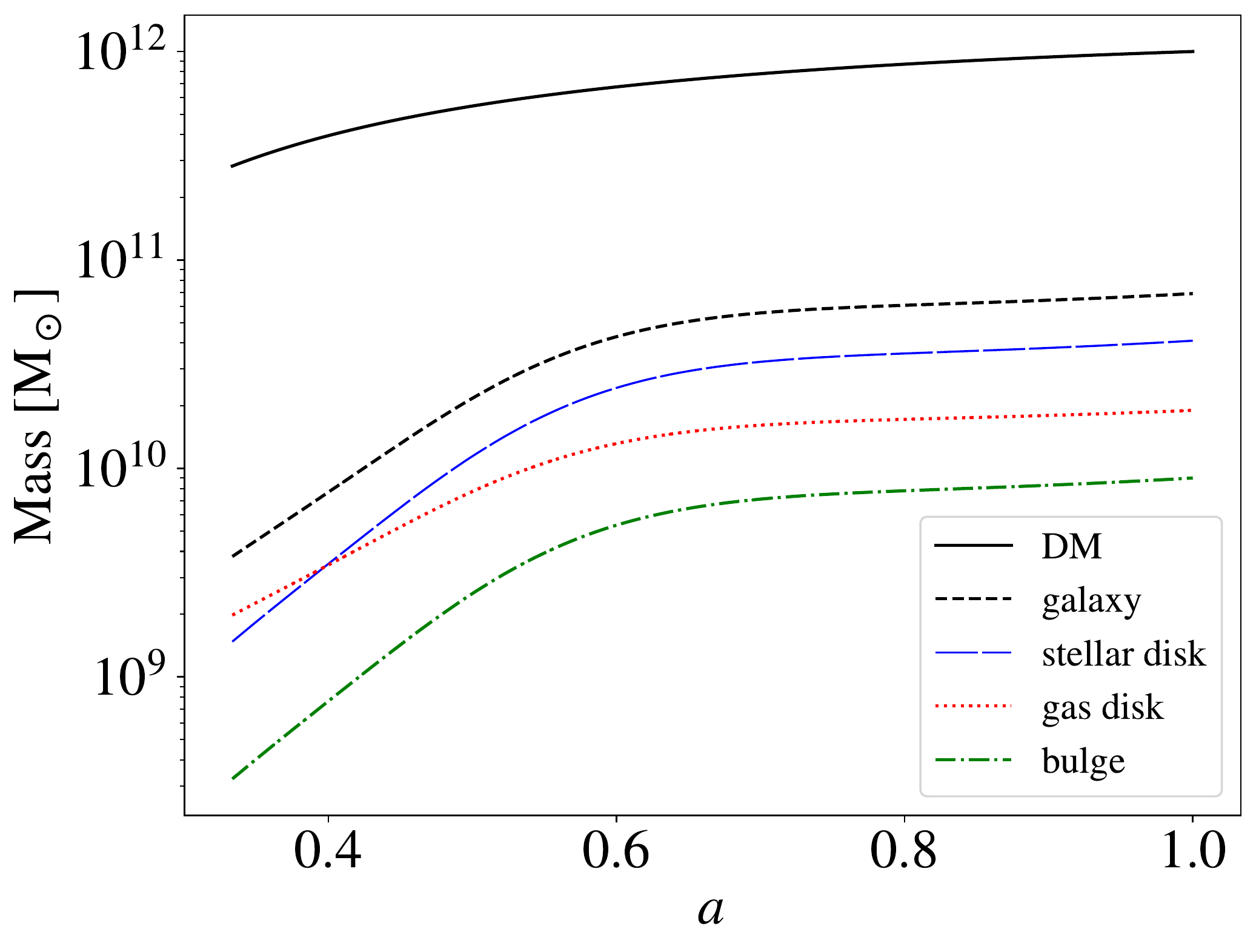}
\\
\includegraphics[width=.97\columnwidth]{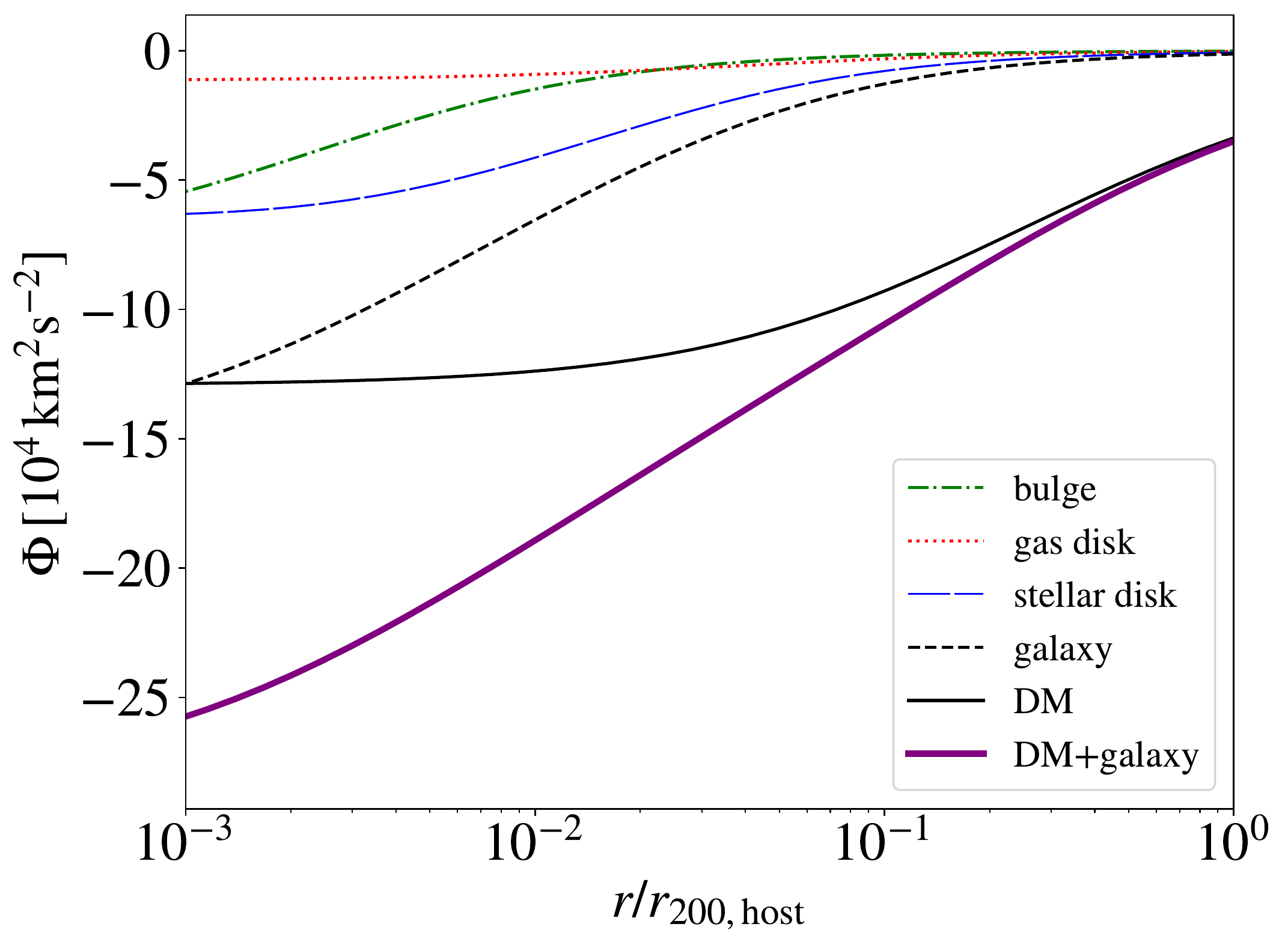}
\caption{Top: Evolution of the various baryon components, similar to~\citet{Kelley2019}. We can see the evolution of the stellar mass in blue, the gas mass in red and the bulge mass in green. The sum of these three is the black solid line and accounts for the total baryonic mass, and the DM mass corresponds to the dashed black line.
Bottom: Values of the potential at $z=0$ for the different components with respect to the radius, normalized to the host virial radius.
}
\label{fig:baryevol}
\end{center}
\end{figure}
The mass of the stellar components, i.e. the stellar disk and bulge, increases with time by following the abundance matching model by~\cite{Behroozi2019}, as the DM host halo mass grows. The mass ratio between the two is fixed as the ratio at $z=0$ (9/41, see \autoref{tab:mwpot}). The gas mass also increases with time, based on the stellar mass and the gas mass fraction by~\cite{Popping2015}. The parameters introducing the spatial scale of the baryon potentials (i.e. the exponential disk scale radius and height, and the bulge scale length) are determined as follows. First, the scale radius of the stellar disk is derived by an empirical relation by~\citet{vanderWel2014}. Then the others are determined to keep the proportion of the baryonic components, e.g. the ratio of the scale radius of the stellar disk to the scale radius of the gas disk is kept. There is another requirement in the model by~\citet{Kelley2019}. The time-varying baryon parameters must be matched by the input parameters at $z=0$. To ensure this requirement, we need three fudge factors\footnote{The evolution of the structural parameters, $M\sub{200,host} (z)$ and $c$, is fully specified as a function of redshift by the empirical relations from cosmological \nbody simulations. Based on the obtained $M\sub{200,host}$  and observationally constrained relations, we can get the expected structural parameters of baryons (mass, scale length and height, etc.) as a function of $z$. However, it is not guaranteed that the derived baryon parameters are consistent with the input parameters at $z=0$. To guarantee the consistency, we need to multiply by the mentioned fudge factors.}, namely $\sim 2$ for the stellar mass, $\sim 0.75$ for the gas mass, and $\sim$ 0.35 for the spatial scale parameters of the baryonic components. They are multiplied by the corresponding parameters. The second fudge factor is within the observed scatter \citep{Popping2015}, 
while the first (third) one seems to be larger (smaller) than the scatter \citep{vanderWel2014,Behroozi2019}. The mass evolution of each host halo baryonic component, as well as its mass in form of DM, is shown in the top panel of Fig.~\ref{fig:baryevol}, while the potential of each component at $z=0$  is shown in its bottom panel.

\subsection{Subhalo orbit} 
\label{ssec:subhalo_orb}

We take only the potential of the spherical DM host halo into account to set the initial subhalo orbit in the host-centric frame (baryon potentials are ignored in setting the initial subhalo orbit). 
An advantage of this scheme is that the initial velocity vector of the subhalo is identical when fixing the orbital parameters. 
The subhalo orbit is characterised with the orbital energy, the angular momentum, and the orbital plane. We employ the following three parameters in this study. The first one describes the orbital energy of the subhalo orbit in the host-centric frame, 
\begin{equation}
    x\sub{c} \equiv r\sub{c}(E)/r\sub{200,host}(z\sub{acc}),
        \label{eq:xc}
\end{equation}
where $r\sub{c}(E)$ and $r\sub{200,host}(z\sub{acc})$ are the radius of a circular orbit of the orbital energy, $E$, and the virial radius of the host halo at the accretion redshift of the subhalo, $z\sub{acc}$, respectively. The second one controls the angular momentum of the orbit,
\begin{equation}
    \eta \equiv L/L\sub{c}(E),
        \label{eq:eta}
\end{equation}
where $L$ and $L\sub{c}(E)$ are the actual angular momentum of the subhalo orbit and the angular momentum of the circular orbit of the energy, $E$. The third parameter is the inclination angle with respect to the galactic plane, $\theta$.

\subsection{Numerical techniques}
\label{ssec:num_tech}

For \nbody computation, we use a code that adopts an oct-tree algorithm~\citep{Barnes1986} and is accelerated with Graphics Processing Units~\citep{Ogiya2013}. The gravitational potential field of particles is smoothed with a~\cite{Plummer1911} force softening of $\varepsilon=0.0003\,r\sub{200,sub}$, where $r\sub{200,sub}$ is the virial radius of the subhalo at accretion. The code employs the cell opening criteria of~\cite{Springel2005_gadget2} with the force accuracy parameter of $\alpha=0.01$. The position and velocity vectors of particles are updated with the second-order Leapfrog scheme 
in each \nbody iteration, and a timestepping is determined with the prescription of~\cite{2003MNRAS.338...14P} 
and is equal for all particles. The centre of the subhalo and its bulk velocity in the host-centric coordinate system is tracked with the scheme outlined in~\cite{vandenBosch2018_analytic}. The evolution of the mass bound to the subhalo is also computed. Only bound particles are considered in drawing the spherically averaged density profile of the subhalo. 

\subsection{Parameter choices}
\label{sec:parchoi}

\begin{figure}
\begin{center}
\includegraphics[width=\columnwidth]{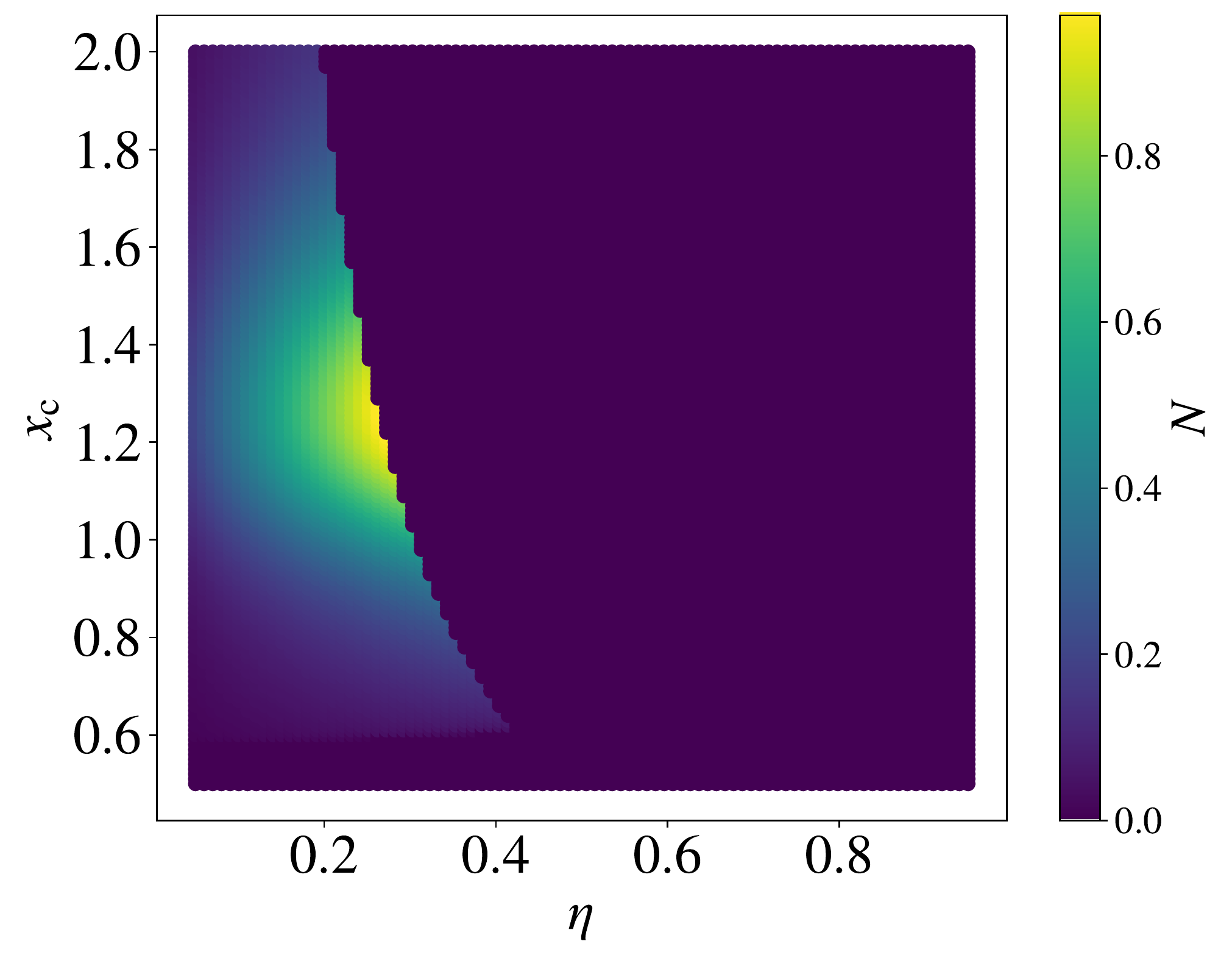}
\caption{Probability distribution function of the orbital parameters of subhaloes, $d^2p/(dx_\mathrm{c}d\eta)$. The fitting function of \citet{2015MNRAS.448.1674J} is used. The evolution of the orbital parameters due to the growth of the MW potential is modelled with the prescription of \citet{Ogiya2021}. Original pairs (i.e. prior to the evolution) to give an orbit crossing the solar galactocentric radius, $R_\odot = 8.5 \ \mathrm{kpc}$, between the accretion redshift, $z_\mathrm{acc} = 2$, and the present time are considered, as such subhaloes are promising targets of DM annihilation signal surveys.
}
\label{fig:pdfrcrit}
\end{center}
\end{figure}

The high numerical accuracy will enable us to study with great detail subhalo survival and its impact in gamma-ray DM searches using the set of parameters that suit best our purposes. We simulate subhaloes with varying mass, concentration, and orbital properties, thus covering the different properties expected in a realistic cosmological scenario. We use six parameters to simulate the subhalo:
\begin{itemize}[leftmargin=0.4cm]
\item[\ding{79}] The initial subhalo mass, $m_\mathrm{sub}$. Since we want to study subhaloes not hosting baryonic material, we have chosen $m_\mathrm{sub} = 10^6~\msun$. We could use even smaller subhalo masses, but this would increase the computational cost significantly so as to cover a much wider dynamical range. In any case, as we will see later, the results both without and with a baryonic host potential are essentially independent on the subhalo mass.
\item[\ding{79}] The subhalo accretion redshift, $z_\mathrm{acc}$. We have chosen $z_\mathrm{acc} = 2$ for most cases since the subhalo accretion distribution in~\citet{2011ApJ...741...13Y} peaks around that value when considering the host and subhalo masses we are working with. It also gives a reasonable amount of subhaloes crossing the solar galactocentric radius at some point along their history, i.e. those expected to be most relevant for DM searches~\citep{Coronado_Blazquez2019}.
\item[\ding{79}] The initial subhalo concentration, $c$. Note that the subhalo is a halo until the moment of accretion, thus the standard definitions of mass and concentration used for haloes are still valid till this happens. As stated in~\citet{2016MNRAS.460.1214L}, the concentration is around 10 for one million solar masses subhaloes (or haloes) being accreted at $z=2$. However, the associated scatter can be considerably larger for smaller subhaloes, so we will study concentration values ranging from 5 to 50. 
This way we would also cover larger concentration values expected for lower mass subhaloes below with $m\sub{sub} = 10^6~\msun$.
\item[\ding{79}] Orbital parameters:
\begin{itemize}[leftmargin=0.4cm]
\item[\ding{99}] The orbital energy parameter, $x_\mathrm{c}$, as described in Section~\ref{ssec:subhalo_orb}. 
\item[\ding{99}] The orbit circularity, $\eta$, as described in Section~\ref{ssec:subhalo_orb}.  
\item[\ding{99}] The orbit inclination angle, $\theta$, as described in Section~\ref{ssec:subhalo_orb}. It is only relevant for runs with baryons, when the host spherical symmetry is broken. 
\end{itemize}
Fig.~\ref{fig:pdfrcrit} presents the probability distribution of $x_\mathrm{c}$ and $\eta$ at the time of subhalo accretion. We employ the fitting function of \citet{2015MNRAS.448.1674J} normalized in the 2D parameter space of $x_\mathrm{c}=[0.5, 2]$ and $\eta=[0, 1]$. As we have a particular interest in subhaloes emitting DM annihilation signals with a detectable flux, subhaloes crossing the solar galactocentric radius, $R_\odot$, at some point since its accretion are considered. After accretion, the subhalo orbit shrinks as a result of the host growth and the pair of the orbital parameters evolves with time. This effect is taken into account by using the model by \citet{Ogiya2021}. We find that subhaloes passing $R_\odot$ typically have $x_\mathrm{c}=1.2$ and $\eta=0.3$ at accretion and adopt this pair as our fiducial choice. 
\end{itemize}

Our choice of orbital parameters, together with the typical initial concentration of haloes at a given redshift, as described above, will constitute what will be called our fiducial set of parameters from now on. Nevertheless, we will vary significantly this fiducial setup in our work by changing the involved parameters to (still reasonable) smaller or larger values, so as to understand the impact of a particular parameter in the results. We summarize both the fiducial setting and the full suite in Table~\ref{tab:fiduset}. 

Finally, we note that our effective mass resolution will depend on the number of particles, $N$. 
In particular, $m_\mathrm{resol} = m_\mathrm{sub} / N $. We choose $N$ such that we try to ensure convergence of results (see later below) 
for the particular set of parameters under consideration within our suite, sometimes increasing it significantly to fulfill this requirement from accretion time to present. Some of the adopted values in this work are listed in Table~\ref{tab:resol}. In this same table, we also show the mean inter-particle distance, $d\sub{ip}$, assuming $N$ particles are homogeneously distributed in a sphere of $r_{200}$. 
Some authors advocate $d\sub{ip}$ sets the minimum value of the softening parameter to ensure the nature of collisionless systems~\citep{1997ApJ...479L..79M, 1998ApJ...497...38S, 2008ApJ...686....1R}. However, in current cosmological $N$-body simulations, the softening is typically larger than some relevant radii, such as the virial radius of a halo resolved with $\sim$ 100 particles, which is $\sim 0.3 d\sub{ip}$~\citep{2022LRCA....8....1A}. The numbers in Table~\ref{tab:resol} show that our mean inter-particle distance fulfills that requirement in most cases, i.e. that it is smaller than the softening length. As expected, the lower the subhalo mass the harder to satisfy the condition.

 \begin{table}
	\centering
	\caption{Set of parameters used in this work, described in Section~\ref{sec:parchoi}. First column: fiducial parameters. Second column: studied range of each parameter in the full suite. 
	}
	\label{tab:fiduset}
	\begin{tabular}{c|cc} 
		\hline
		  & fiducial & suite  \\
		  \hline
		  $m_\mathrm{sub} [\mathrm{M}_\odot]$ & $10^6$ & $[1,10^9]$ \\
		  $z_\mathrm{acc}$ & $2$ & $[1,4]$ \\
		  $c$ & $10$ & $[5,50]$ \\
		  $ \eta$ & $0.3$ & $[0.1,0.8]$ \\
		  $x_\mathrm{c}$ & $1.2$ & $[0.8,1.6]$ \\
		  $\theta $ [deg] & 45 & $[0,90]$ \\
		\hline 
	\end{tabular}
\end{table}
 
 \begin{table}
	\centering
	\caption{Examples of mass resolution, mean inter-particle distance and softening length for 
	some subhalo masses and number of particles used in this work. All cases assume $z_\mathrm{acc} = 2$.}
	\label{tab:resol}
	\begin{tabular}{ccccc} 
		\hline
		  $m_\mathrm{sub} [\mathrm{M}_\odot]$ & $N$ & $m_\mathrm{resol} [\mathrm{M}_\odot]$ & $d\sub{ip}$ [pc] & $\varepsilon$ [pc]  \\
		\hline 
		$10^6$ & $2^{18}$ & $3.81$ & $0.16$ & $0.304$ \\
		$10^6$ & $2^{20}$ & $0.95$ & $0.099$ & $0.304$ \\
		$10^6$ & $2^{21}$ & $0.48$ & $0.078$ & $0.304$ \\
		$10^3$ & $2^{20}$ & $0.00095$ & $0.046$ & $0.0304$ \\
		$1$  & $2^{21}$ & $4.77 \cdot 10^{-7}$ & $0.017$ & $0.00304$ \\
		\hline
	\end{tabular}
\end{table}

\section{Results}\label{sec:resul}

In this section, we summarize the main findings in our analyses. We have mainly studied two relevant quantities: the bound mass fraction, $f_\mathrm{b}$, which corresponds to the fraction of the initial subhalo mass that remains bound at a given redshift, and the annihilation luminosity, $L$, which is defined as the integral of the subhalo density profile squared.

\subsection{Bound mass fraction}

The bound mass fraction comprises the information about how much mass the subhalo has lost when a certain amount of time has passed since its accretion. 
We define it as the fraction of mass that remains bound at time $t$ with respect to the initial subhalo mass~\citep{vandenBosch2018_analytic}: 

\begin{equation}
    f_\mathrm{b} = \frac{M(t)}{M_\mathrm{200,sub}},
\end{equation}

where $M(t)$ is the bound mass of the subhalo at time $t$, and $M_\mathrm{200,sub}$ = $M(< r_\mathrm{200,sub})  = m_\mathrm{sub}$ is the initial virial mass of the subhalo. This virial radius will not be a good parameter to define the subhalo after accretion, since the mass at the outskirts will be eventually lost and its profile will be consequently truncated.

This quantity allows us to elucidate if the subhalo has been disrupted or if it survives after several orbits. We study $f_\mathrm{b}$ for the cases in which the host is made of DM alone as well as the one in which baryons are also included following our prescription in Section~\ref{sssec:mw}. These cases are detailed, respectively, in the next Sections~\ref{sec:fbnonbary} and~\ref{sec:fbbary}. 
Furthermore, we study the values of  $f_\mathrm{b}$ that can be trusted in our analyses via the definition of strict convergence criteria in either case, which are nailed down in Appendix~\ref{sec:apfb} for the interested reader.

\subsubsection{Non-baryonic case}\label{sec:fbnonbary}

\begin{figure}
\begin{center}
\includegraphics[width=\columnwidth]{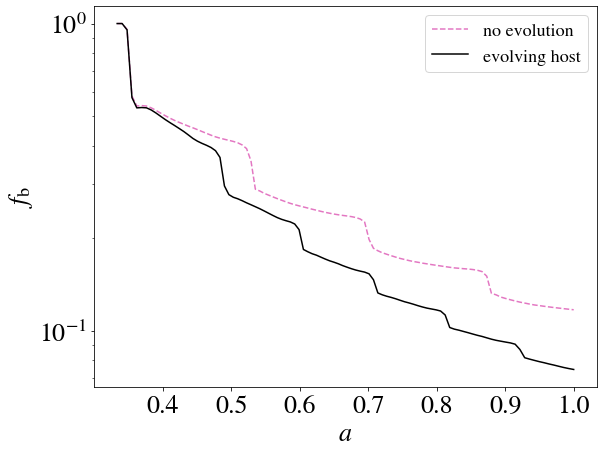}
\caption{ Bound mass fraction as a function of $a = 1/(1+z)$ for a one-million-solar-mass subhalo orbiting a host halo without baryons. We compare two cases: the solid black line corresponds to a scenario in which the DM host potential evolves, while the dashed pink line is for a static host with no evolution.  
 The static mass and concentration are $M_\mathrm{200,host} = 2.82 \cdot 10^{11} \mathrm{M_\odot}$ and $c_\mathrm{host} = 5.14$, respectively, which are the evolving case initial values, i.e. they are fixed as those at $z=z_\mathrm{acc}$. 
In both cases we have used our fiducial set of parameters given in Table~\ref{tab:fiduset}.
} 
\label{fig:fbhev}
\end{center}
\end{figure}

First, we study the effect that the time evolution of the DM host potential has in the mass loss process. This is a new feature of our code, not included in DASH nor shown before. The difference between including this effect or not is illustrated in Fig.~\ref{fig:fbhev} for a particular example. A larger subhalo depletion as well as a larger number of pericentric passages are observed in this more realistic scenario. Most significant changes occur at the pericentre, when a larger fraction of material from the subhalo is stripped by the host (appearing as abrupt `steps' in this figure). 
In this particular case, the subhalo whose host evolves loses more mass mainly because it experiences a higher number of pericentric passages. The apocentre is smaller and decreases with time as well. We have checked different cases finding essentially the same results. From now on, we will always adopt the case of an evolving host as the fiducial one, unless specified otherwise. 

\begin{figure*}
\begin{center}
\includegraphics[width=.495\textwidth]{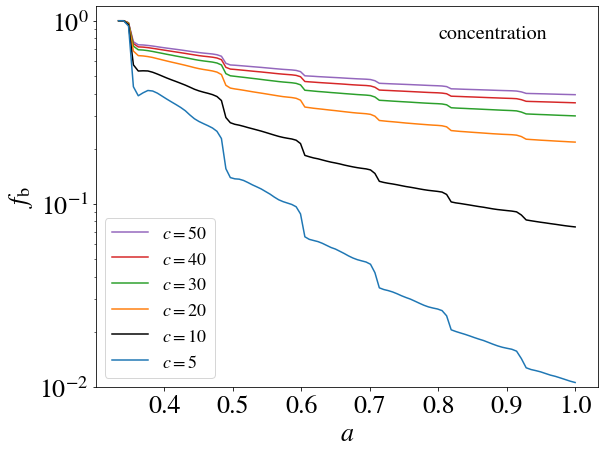}
\includegraphics[width=.495\textwidth]{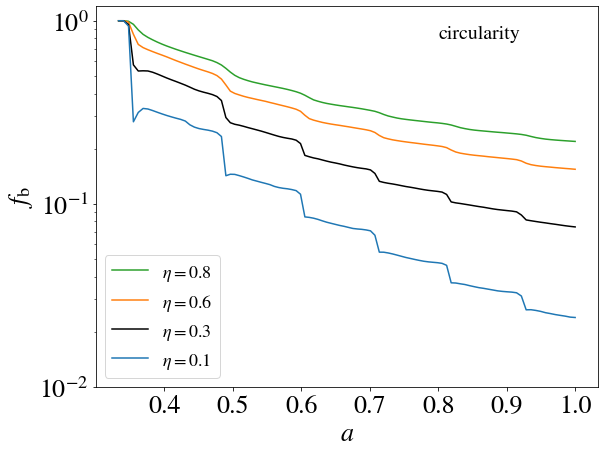}
\includegraphics[width=.495\textwidth]{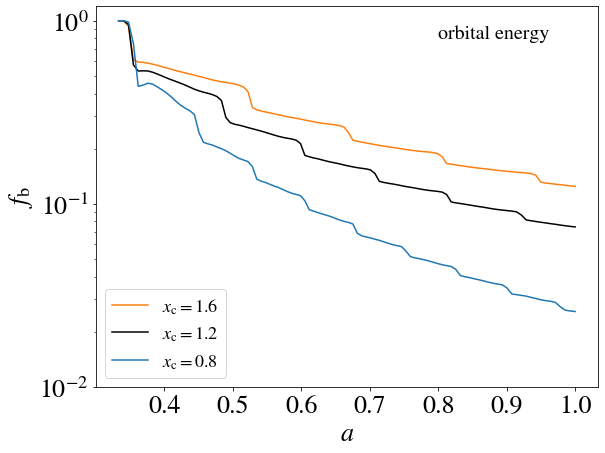}
\includegraphics[width=.495\textwidth]{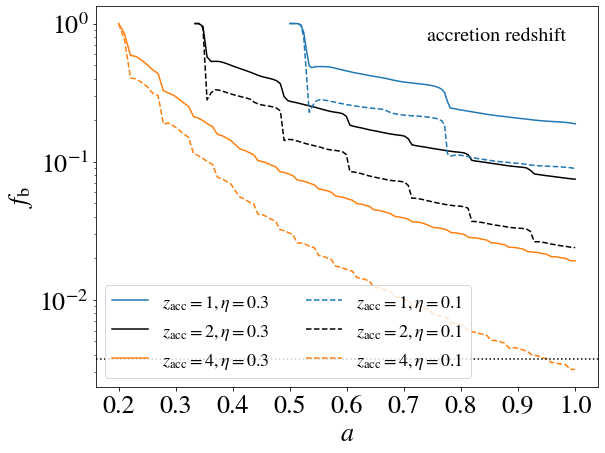}
\caption{Bound mass fraction, $f_\mathrm{b}$, as a function of $a = 1/(1+z)$ for different subhalo configurations. Each of them corresponds to a case in which we vary one parameter among those representing our fiducial setup in Table~\ref{tab:fiduset}. The latter is depicted as a solid black line in all panels for reference. 
Upper left panel: Different initial subhalo concentrations ($c$).
Upper right panel: Different initial circularities ($\eta$).
Lower left panel: Different orbital energies ($x_\mathrm{c}$).
Lower right panel: Different accretion redshifts ($z_\mathrm{acc}$) using two different circularity values.
The black horizontal dotted line sets the convergence value for $f_\mathrm{b}$, explained in Appendix~\ref{sec:apfb}. 
When it does not appear, this value is below the chosen $y$-axis lower limit. Note that $f_\mathrm{b}$ is always well above except in the lower right panel, for which no convergence is achieved at present time for the case of $z_\mathrm{acc} = 4, \eta = 0.1$.
} 
\label{fig:fb1}
\end{center}
\end{figure*}

 In 
 Fig.~\ref{fig:fb1}, we show $f_\mathrm{b}$ as a function of the scale factor, $a = 1/(1+z)$, for different subhalo configurations. In each of them, we vary a parameter among those defining our fiducial setup specified in Table~\ref{tab:fiduset}. 
 In particular, in the upper panels of  Fig.~\ref{fig:fb1}, we show the evolution of $f_\mathrm{b}$ for different concentrations and circularities, respectively. From these panels one can see that less concentrated subhaloes at accretion lose mass more quickly, which agrees with the expectations. 
 Also, more radial orbits, i.e. those with smaller $\eta$, imply a larger mass loss. Note that we are comparing different eccentricities here while fixing $x_\mathrm{c}$. Therefore, our orbits with higher eccentricities have smaller pericentres and experience a stronger tidal field. In the lower left panel of Fig.~\ref{fig:fb1} different orbital energy parameter values are displayed. In this case, a smaller $x_\mathrm{c}$ leads to a larger number of orbits in the same time interval and, thus, to a greater mass loss as well. 
Finally, the lower right panel shows examples for different accretion redshifts, and we can see that a larger $z_\mathrm{acc}$ has also the effect of inducing more mass loss: the subhalo completed more orbits and it initially had a smaller pericentre because the host halo was smaller at earlier cosmic epochs. Indeed, subhaloes accreted at different times landed on different orbits and later-accreted subhaloes have spent less time within the host. We use $N = 2^{18}$ particles in most cases, increasing this number up to $N = 2^{21}$ whenever needed.

A general picture of $f_\mathrm{b}$ results at $z=0$ in the non-baryonic case can be seen in the upper left panel of Fig.~\ref{fig:fbsurf}. In this plot, we fix $x_\mathrm{c} = 1.2$ and $z_\mathrm{acc} = 2$ and vary both the concentration and $\eta$ parameters. The summary is that subhaloes lose less mass when any of these two parameters is larger. These results are expected to be scale-free when the subhalo mass is small enough. More specifically, results will be identical for ratios $M_\mathrm{200,host}/m_\mathrm{sub} \gtrsim 10^3$, since self-friction becomes negligible~\citep{Ogiya2019,2020MNRAS.495.4496M}. 
Actually, dynamical friction would work more significantly than self-friction in decaying subhalo orbits \citep{2020MNRAS.495.4496M}. Nevertheless, when considering subhaloes of low enough masses, this drag force would be negligible as well.

\subsubsection{Baryonic case}\label{sec:fbbary}

Recent hydrodynamical simulations have shown a significant decrease of the number of subhaloes when baryons are taken into account~\citep{2017MNRAS.471.1709G, Kelley2019, 2021MNRAS.501.3558G, 2021MNRAS.tmp.2848G}. 
However, this could still be due to numerical artifacts related to insufficient mass and/or force resolution~\citep{vandenBosch2018_analytic, vandenBosch2018_num_criteria}. 
In our work, we want to give an answer to this ongoing debate by performing a variety of simulations including the baryonic component of a MW-size halo as well.
Therefore, in order to obtain more realistic simulations we now add baryonic material to the host potential. This has been done 
as described in Section~\ref{sssec:mw}, in a way such that the baryonic analytical potential also evolves with time, from accretion until the present. %

\begin{figure*}
\begin{center}
\includegraphics[width=.495\textwidth]{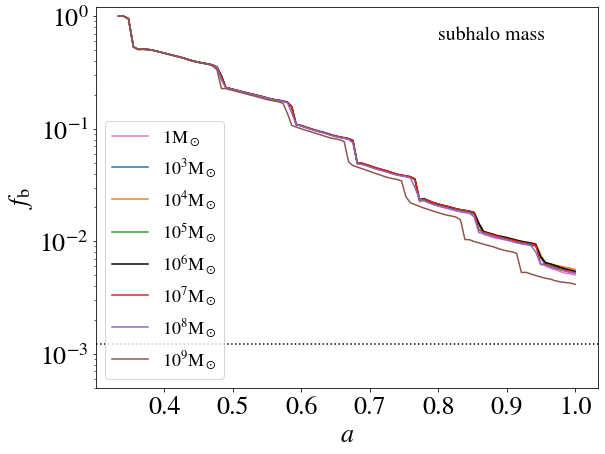} 
\includegraphics[width=.495\textwidth]{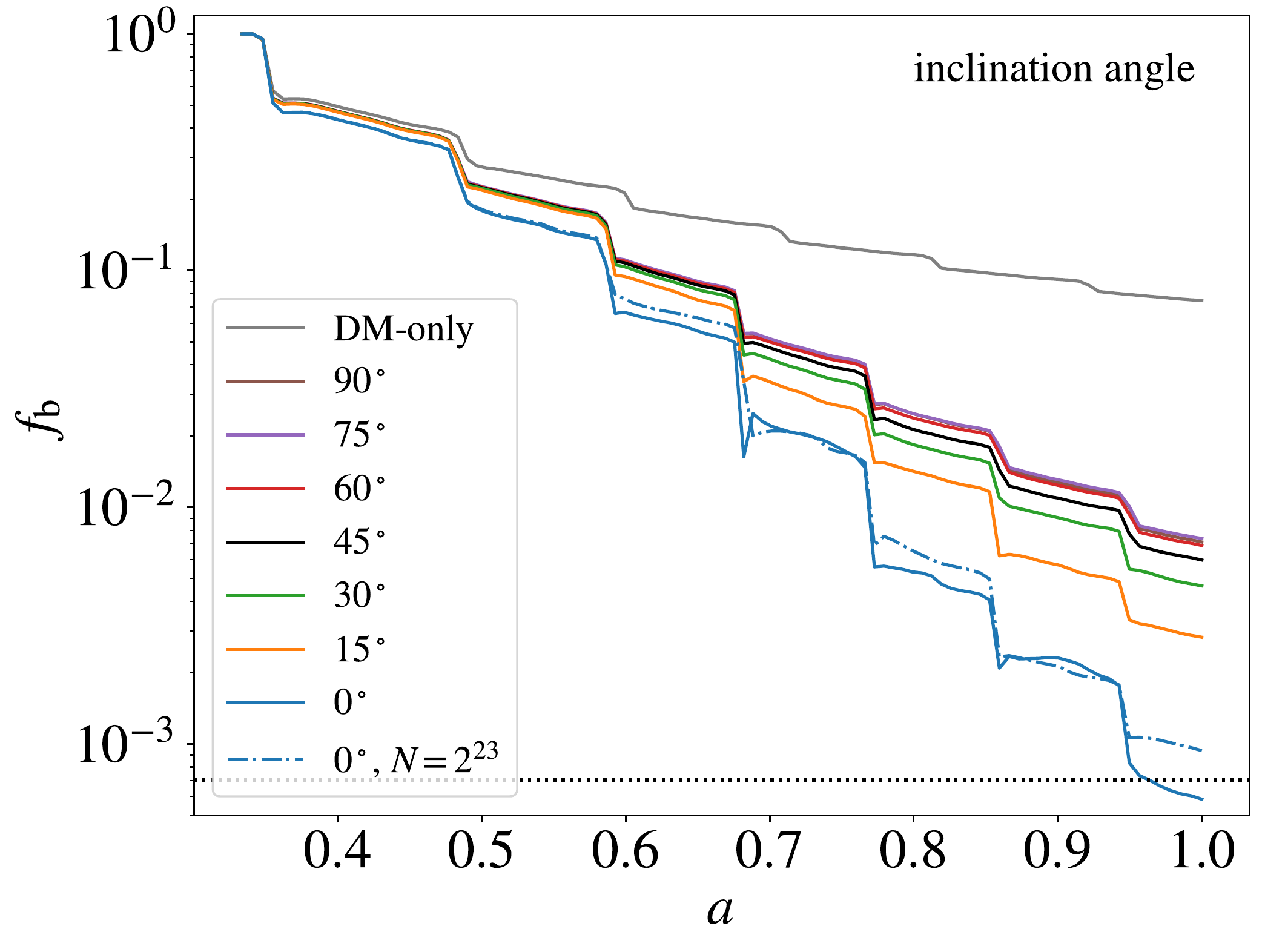} \\
\caption{Checks including baryons. 
Left: 
Bound mass fraction, $f_\mathrm{b}$, as a function of $a = 1/(1+z)$ for a subhalo described with our fiducial set of parameters (Table~\ref{tab:fiduset}), but different initial mass and orbiting a MW-like host with baryonic material. In all cases, the orbit inclination angle with respect to the galactic disk is 45 degrees. 
Right:  
Same like in the left panel, this time for a one-million-solar-mass subhalo orbiting its host with different inclination angles as given in the legend. Zero degrees means that the orbit is parallel to the disk, while 90 represents a perpendicular orbit. Also shown for comparison are the lines corresponding to the non-baryonic case (grey line) and a run with 4 times higher resolution (dot-dashed line). In both panels, the horizontal dotted lines represent the convergence limit (for the $N=2^{20}$ and $N=2^{21}$ resolution cases in the left and right panels, respectively; see Appendix~\ref{sec:apconv} for details). 
}
\label{fig:fullbary}
\end{center}
\end{figure*}

First of all, we want to understand if also in the case of including baryons the results are scale-free when the subhalo mass is small enough. This was only shown before for the DM-only case \citep[][see also \citealt{2022arXiv220700604S}]{Ogiya2019}. 
Our findings are depicted in the left panel of Fig.~\ref{fig:fullbary}. 
We can see that results are very similar for masses ranging from one solar mass up to ten million solar masses. 
We have also checked the impact of self-friction~\citep{2020MNRAS.495.4496M} for larger masses. Indeed, this effect starts to be noticeable at $10^8\, \mathrm{M_\odot}$ --the orbits become smaller, which leads to more mass loss-- and it is significant for $10^9\, \mathrm{M_\odot}$ subhaloes, the difference in $f_\mathrm{b}$ at $z=0$ being a factor $\sim$1.5. In contrast, \citet{Ogiya2019} found no noticeable difference up to $10^9\, \mathrm{M_\odot}$ for runs without baryons. 

We have also investigated the influence of the orbit inclination angle in subhalo depletion, also studied in~\citet{2021MNRAS.tmp.2848G}. Similarly to the latter work, our results, depicted in the right panel of Fig.~\ref{fig:fullbary}, show that subhaloes with orbits more parallel to the baryonic disk lose more mass. Yet, we report more substantial mass loss for parallel orbits.\footnote{We have checked that using a larger concentration, $c=20$, as they do, diminishes this difference significantly. We also note that their host potential consists of an NFW DM halo and a single Miyamoto-Nagai disk to account for baryons, both being static, while ours is more elaborated as described in Section~\ref{ssec:host_pot}.} 
Other works~\citep{2010ApJ...709.1138D, 2017MNRAS.471.1709G} suggested that the mass loss for a more perpendicular orbit would be greater due to disk shocking when the subhalo suddenly enters or leaves the baryonic tidal field. Here we find the relevance of this potential effect to be negligible. 
Instead, we found that another parameter, namely the force accuracy, becomes particularly relevant for parallel orbit runs, since the subhalo in these orbits can deviate from the disk plane after several pericentric passages if the force accuracy is not good enough, which, in turn, causes a small difference in $f_\mathrm{b}$.

\begin{figure}
\begin{center}
\includegraphics[width=\columnwidth]{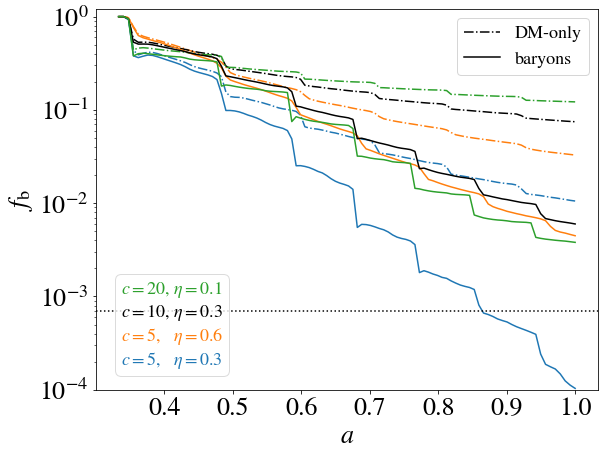}
\caption{Comparison between adding baryons to the host halo potential or having only DM, for both different concentration and orbital circularity values. The fiducial setup of Table~\ref{tab:fiduset} is shown as a black line. Non-baryonic runs are depicted as dash-dotted lines while baryonic ones are in solid. The black dotted horizontal line corresponds to the convergence limit; see Appendix~\ref{sec:apconv}. In all cases, we set $x_\mathrm{c} = 1.2$, $z_\mathrm{acc} = 2$ and 
$m_\mathrm{sub} = 10^6 \mathrm{M_\odot}$. When baryons are included, the inclination angle is 45 deg. 
}
\label{fig:compbary}
\end{center}
\end{figure}

In Fig.~\ref{fig:compbary}, we compare the impact that adding baryons or not to the host potential has on the bound mass fraction.
We adopt an inclination angle of 45 degrees in this example as an intermediate choice.
As it can be seen, the presence of baryonic material can have a huge impact on the subhalo depletion, especially when the pericentre of the orbit is smaller (e.g. decreasing $\eta$ while fixing $x_\mathrm{c}$). This typically leads to a much smaller $f_\mathrm{b}$ for the same time after accretion when compared to the non-baryonic case. Indeed, Fig.~\ref{fig:compbary} shows that some non-baryonic runs with smaller $\eta$ but larger $c$ can lead to comparatively less mass loss, while this is not necessarily the case when baryons are included.

\begin{figure*}
\begin{center}
\textbf{Bound mass fraction \ \ \ \ \ \ \ \ \ \ \ \ \ \ \ \ \ \ \ \ \ \ \ \ \ \ \ \ \ \ \ \ \ \ \ \ \ \ \ \ \ \ \ \ \ \ \ \ \ \ \ \ \ \ \ \ \ \ \ \ \ \ \ \ \ \ \ \ \ \ \ \ \ \ \ \ \ \ \ \ \ \ \ \ \ \ \ \ \ \ \ \ \  Annihilation luminosity}\par\medskip
\vspace{-0.2cm}
\includegraphics[width=.5\textwidth]{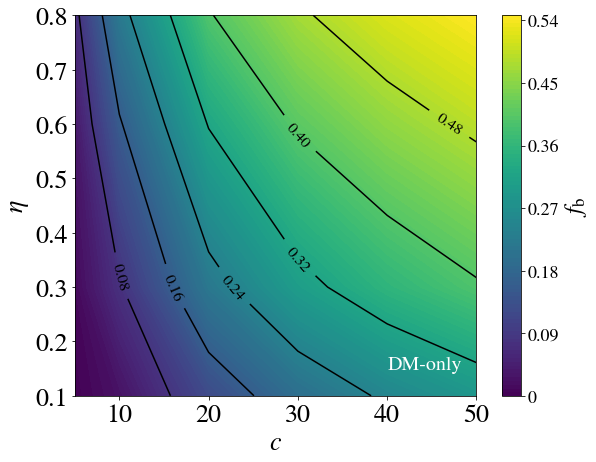}
\includegraphics[width=.49\textwidth]{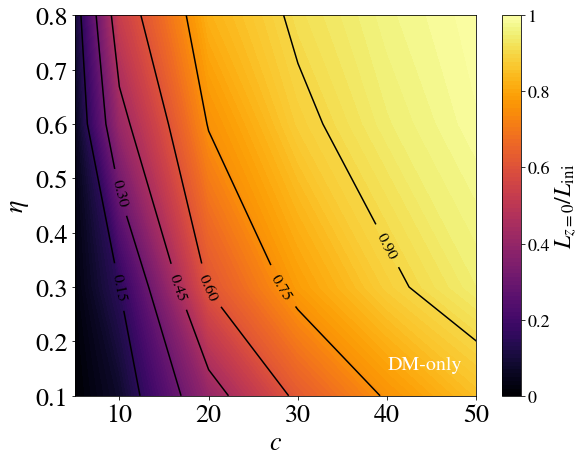}
\includegraphics[width=.5\textwidth]{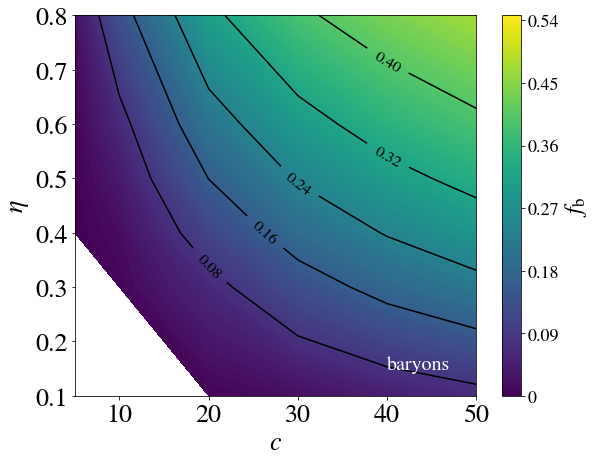}
\includegraphics[width=.49\textwidth]{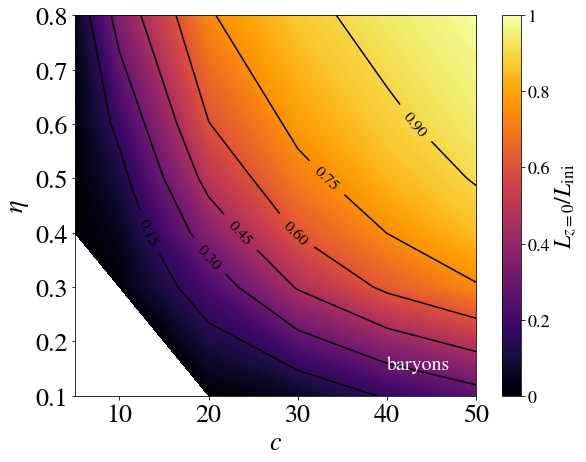}
\includegraphics[width=.5\textwidth]{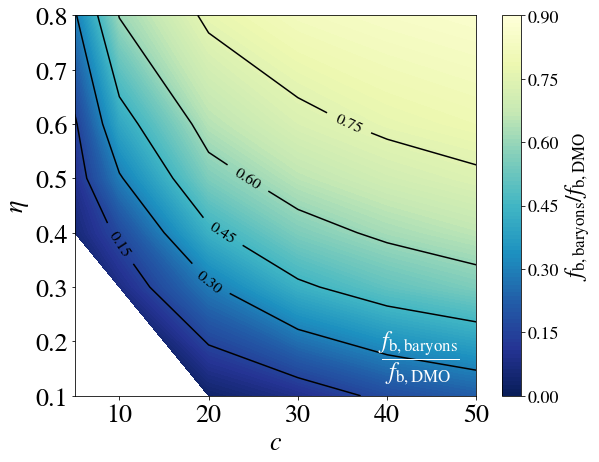}
\includegraphics[width=.49\textwidth]{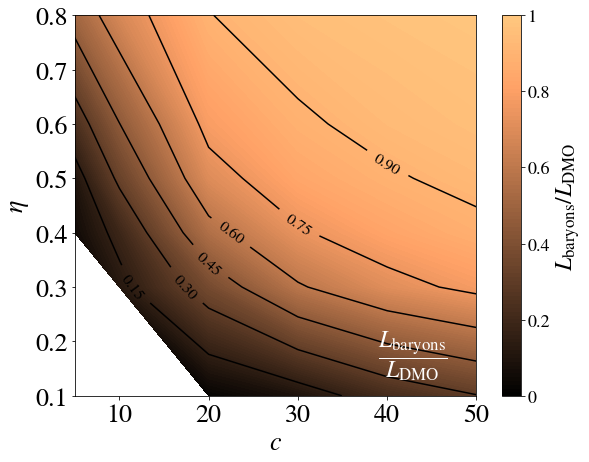}
\caption{Left: Bound mass fraction, $f_\mathrm{b}$, at present time for different initial subhalo concentrations and circularities in the non-baryonic case (top) and in case of including baryons (middle). 
The ratio between the two previous panels, i.e. $ f_\mathrm{b,baryons}/f_\mathrm{b,DMO}$ at $z=0$, is shown in the bottom panel.
Right: Annihilation luminosity results at $z=0$ varying both the concentration and $\eta$ parameters, both for the case of excluding baryons (top panel) and with baryons included (middle). The bottom panel shows the ratio between the two previous panels, i.e. $L_\mathrm{baryons}/L_\mathrm{DMO}$ at $z=0$.
We adopt a one-million-solar-mass subhalo, with $x_\mathrm{c} = 1.2$ and $z_\mathrm{acc} = 2$ in all cases, and fix the inclination angle of the subhalo orbit to 45 degrees in the case of baryons. Non-converged runs are not shown and are the cause of the blank regions in these plots.
} 
\label{fig:fbsurf}
\end{center}
\end{figure*}

A general picture of $f_\mathrm{b}$ results at $z=0$ for the runs including baryons can be seen in the middle left panel of Fig.~\ref{fig:fbsurf}. In this plot, we fix the inclination angle of the subhalo orbit to 45 degrees, adopt $x_\mathrm{c} = 1.2$ and $z_\mathrm{acc} = 2$, and vary both the concentration and $\eta$ parameters. Again, we conclude that subhaloes lose less mass when any of these two last parameters is larger. We note that we cannot achieve numerical convergence for a few cases in our grid\footnote{We have tried to improve the convergence using different values of $N$, up to $2^{21}$, but did not succeed. We note though that enlarging $N$ even more drastically should allow to reach a convergent run in the end for most cases; however, the computational resources needed to do so were too expensive.}, although we do for most of them. An example of the latter can be actually seen as the blue solid line in Fig.~\ref{fig:compbary} as well. 
The lower left panel of Fig.~\ref{fig:fbsurf} shows the ratio between baryonic and DM-only runs, and confirms again the larger impact of baryons, especially for subhaloes in more radial orbits. We find the largest differences in the lower left area for the lowest $c$ and $\eta$ values considered. Besides, on the bottom right corner, where $c$ is large but $\eta$ is small, this ratio reaches values $\sim 0.25$, while the ratio is $\sim 0.3$ on the upper left corner. When both $c$ and $\eta$ are large (upper right area), both values are similar.

\subsection{DM annihilation luminosity}\label{sec:lumi}

Studying the annihilation luminosity of galactic subhaloes is essential to understand their potential as targets for gamma ray searches~\citep{2012ApJ...747..121A, 2013PhR...531....1S, 2017JCAP...04..018H}. For instance, current DM constraints obtained from the scrutiny of unidentified gamma-ray sources in search of potential subhaloes with no visible counterparts depend, in the first place, on the number of detectable subhaloes predicted from a combination of simulations and instrumental sensitivity~\citep{Coronado_Blazquez2019, 2021PDU....3200845C, 2022PhRvD.105h3006C}. More specifically, these DM constraints would be overly optimistic if a significant fraction of subhaloes in the solar vicinity disrupt or lose a significant fraction of their luminosity. Having more resilient subhaloes than those in current simulations would also impact the mentioned DM constraints, this time in the opposite way. Thus, for these studies it is important to have robust predictions of the number of subhaloes, probably down to scales as low as one thousand solar masses~\citep{Coronado_Blazquez2019}. In particular, knowing both the precise abundance and radial distribution of the subhalo population within a MW-like host would be of utmost importance, not only from a purely cosmological perspective and for current DM constraints, but also e.g. to understand the role of subhaloes for the so-called subhalo annihilation boost~\citep{mascprada14, Moline17, Ando:2019xlm, SpecialIssueGal}

The way to compute the subhalo luminosity is via the radial density profile $\rho(r)$; more specifically, we define the annihilation luminosity in our study as the integration of the DM density profile squared: 
$L = \int_{V}  \rho^{2}(r) \, dV$.\footnote{Note that this actually corresponds to the annihilation luminosity in the case of a velocity-independent annihilation cross section. If a velocity dependence was included, additional factors may come in.}
The fraction of this annihilation luminosity that reaches the Earth and we can potentially measure with our telescopes is the annihilation flux. 
We note, however, that the latter cannot be predicted without knowing the exact distance between the subhalo and us. 

\begin{figure}
\begin{center}
\includegraphics[width=\columnwidth]{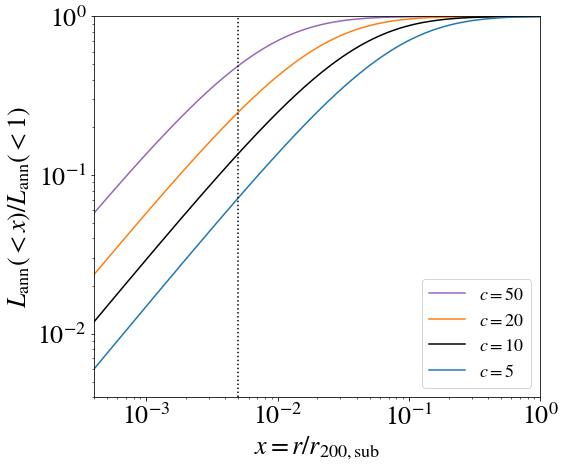}
\caption{Fraction of the total annihilation luminosity in the region contained inside a radius $r$, normalized to the subhalo virial radius, for different initial subhalo concentrations. The black dotted line corresponds to the relaxation radius for $N = 2^{24}$ particles, as indicated in Section~\ref{sec:lumi} and explained in Appendix~\ref{sec:aplu}.
} 
\label{fig:anlumisins}
\end{center}
\end{figure}

The fraction of the annihilation luminosity contained inside a normalized (sub)halo radius $x$, adopting an NFW DM density profile, is shown in Fig.~\ref{fig:anlumisins} for different initial subhalo concentrations. It can be seen that, even for small concentration values, more than 10\% of the total luminosity is inside 1\% of the virial radius, and can be more than half for larger concentrations.  Therefore, the lack of numerical resolution in the innermost part of the subhalo together with the effect of particle relaxation (explained in detail in Appendix~\ref{sec:aplu}) makes the study of the annihilation luminosity a difficult task. Indeed, no particle data are available inside 0.1\% of the initial virial radius of the subhalo, and we lose some 
of these inner particles --up to 1-3\% of the initial virial radius, depending on $N$-- after several pericentric passages.\footnote{The fraction of lost particles within the innermost 1\% of the subhalo initial radius depends on the specs of the particular run, reaching up to $\sim$80\% in some cases.} 
This happens specially when the initial concentration is small. 
To shed further light on this potential issue for annihilation luminosity, we have analyzed the change in the density profile of a subhalo as the number of particles increases, and found that the inner cusp remains when $N$ is large enough ($N \gtrsim 2^{22}$; see Fig.~\ref{fig:aplumis}), but becomes a core when it is not.  This implies a significant, non-physical luminosity loss. To solve this problem, we reconstruct the inner cusp in each snapshot in a semi-analytical way. Full details of this cusp reconstruction can be found in Appendix~\ref{sec:aplu}.

\begin{figure*}
\begin{center}
\includegraphics[width=.495\textwidth]{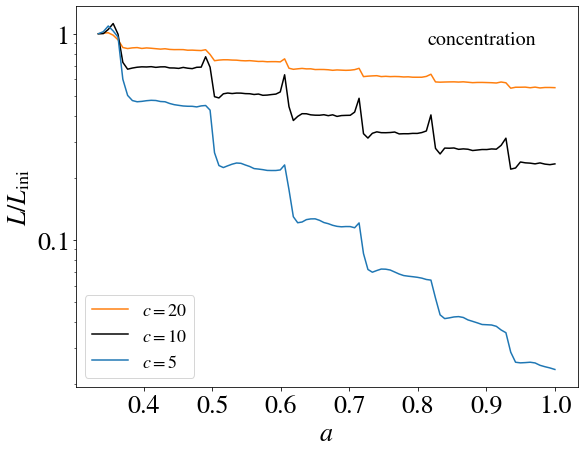}
\includegraphics[width=.495\textwidth]{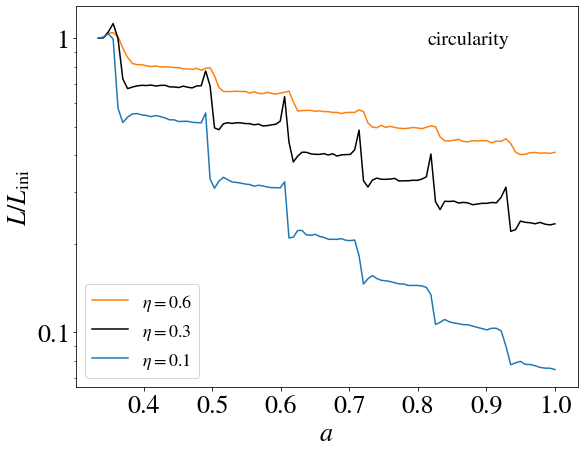} 
\includegraphics[width=.495\textwidth]{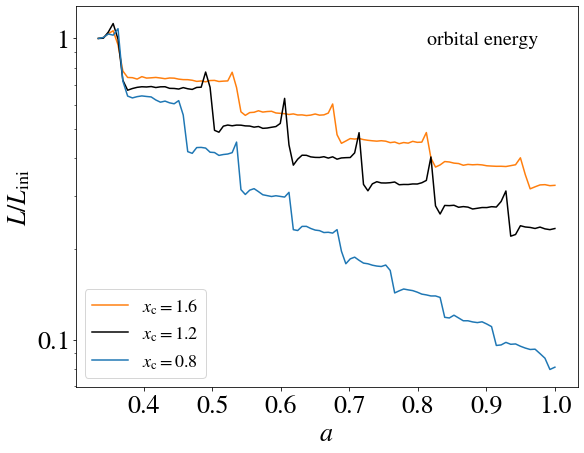}
\includegraphics[width=.495\textwidth]{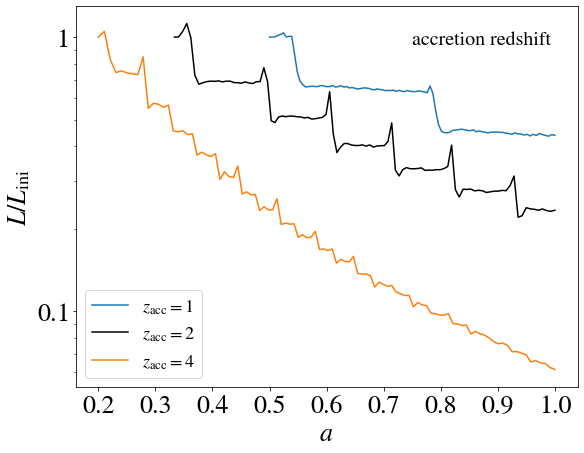}
\includegraphics[width=.495\textwidth]{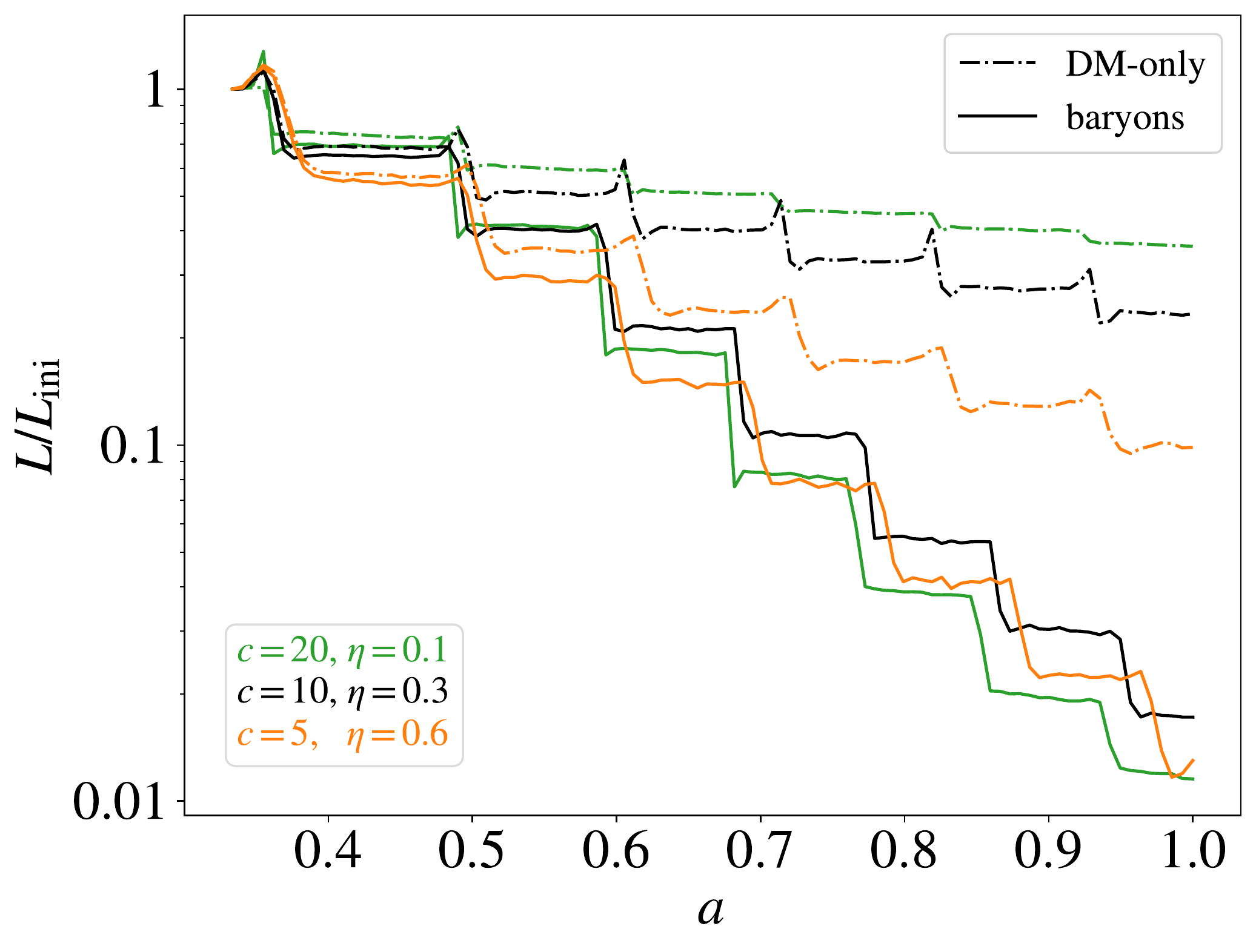}
\includegraphics[width=.495\textwidth]{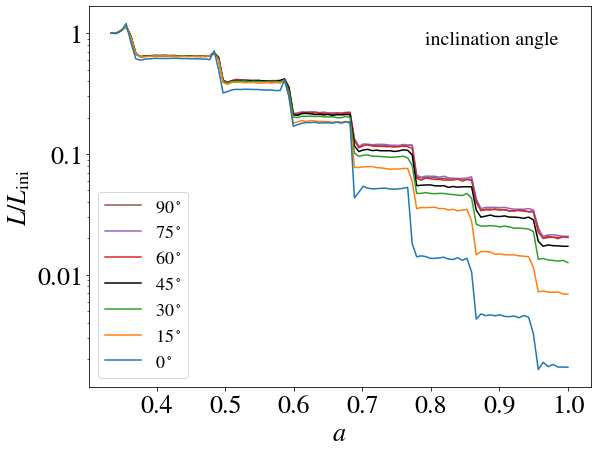}
\caption{Evolution of the annihilation luminosity normalized to its initial value at accretion, $L/L_\mathrm{ini}$, as a function of the scale factor, $a = 1/(1+z)$, for different subhalo configurations. In each panel (except for the lower left), we vary one single parameter with respect to the fiducial setup in Table~\ref{tab:fiduset}. The latter corresponds to the black solid line in all panels. 
Comparison for different initial subhalo concentrations, $c$ (upper left); different initial circularities, $\eta$ (upper right); orbital energies, $x_\mathrm{c}$ (middle left); accretion redshifts, $z_\mathrm{acc}$ (middle right); comparison for runs with and without baryons (lower left); 
and different orbit inclination angles for the baryonic case (lower right). 
}
\label{fig:lumievo}
\end{center}
\end{figure*}

In Fig.~\ref{fig:lumievo}, we show the evolution of the annihilation luminosity normalized to its initial value at accretion, $L/L_\mathrm{ini}$, as a function of the scale factor, $a = 1/(1+z)$, for different subhalo configurations. In each panel of this figure, we vary one single parameter with respect to the fiducial setup of Table~\ref{tab:fiduset}.
 The first four panels show runs without baryons.
 In particular, in the upper panels we show the evolution for different concentrations and circularities, respectively. We conclude that less concentrated subhaloes at accretion 
 get reduced to a smaller fraction of their initial luminosity (by e.g. a factor $\sim 4$ in the fiducial case), which is in tune with expectations. 
 Also, more radial orbits, i.e. those with smaller $\eta$, experience the same effect. Note that we are comparing different eccentricities here while fixing $x_\mathrm{c}$. Therefore, our orbits with higher eccentricities have smaller pericentres and undergo a stronger tidal field. 
 In the middle left panel, different orbital energy parameters are displayed. We observe that smaller $x_\mathrm{c}$ values lead to a larger number of orbits in the same time interval and a larger luminosity decrease as well. 
The middle right panel shows results for different accretion redshifts: larger $z_\mathrm{acc}$ also allows for more orbits and a smaller pericentre (because the host halo was smaller at younger cosmic epochs) and, thus, the luminosity is significantly lower at present for earlier accreted subhaloes. 
 A comparison between runs without and with baryons is shown in the lower left panel. Notice again that $\eta$ becomes relevant when baryons are included, since a small value induces a greater change in the luminosity. 
 Lastly, the lower right panel shows the luminosity for orbits with different inclination angles with respect to the baryonic disk, confirming that subhaloes in parallel orbits become less luminous after several pericentric passages.

A general picture of annihilation luminosity results at $z=0$ varying both the concentration and $\eta$ parameters can be seen in Fig.~\ref{fig:fbsurf}, both for the case of excluding baryons (top right panel) and with baryons included (middle right). We adopt $x_\mathrm{c} = 1.2$ and $z_\mathrm{acc} = 2$ in all cases, and fix the inclination angle of the subhalo orbit to 45 degrees in the case of baryons.
As in the case of $f_\mathrm{b}$, we do not reach numerical convergence for a few cases in our grid, although we do for most of them.
For our fiducial subhaloes (Table~\ref{tab:fiduset}) there is always a significant reduction of luminosity, the subhalo retaining about 15\% and 2\% of its initial luminosity in the non-baryonic and baryonic cases, respectively.
More in general, it can be seen that the concentration is the most relevant parameter when baryons are not considered, the subhalo not losing a significant luminosity fraction when $c$ is large enough, while also $\eta$ plays a major role when baryons are added to the game. 
More specifically, baryons have a large impact on the annihilation luminosity when the orbits are more radial (smaller $\eta$) since the subhalo gets closer to the host halo centre, where baryons are mostly located, thus enhancing the disruption. 
This is more clearly visible in the bottom right panel of the same Fig.~\ref{fig:fbsurf}, which shows the ratio between annihilation luminosities found at $z=0$ in the baryonic and DM-only cases. The largest differences are located in the lower left area for the lowest $c$ and $\eta$ values considered. But still on the bottom right corner, where $c$ is large but $\eta$ is small, this ratio reaches values $\sim 0.3$. In contrast, the ratio for $c = 5$ and $\eta = 0.8$ is $\sim 0.5$ (upper left corner of the plot). When both $c$ and $\eta$ are large (upper right), both values are similar.

\section{Discussion}\label{sec:discu}


In this section we try to simplify the parameter space of tidal stripping by summarizing it into a single parameter. First, in Section~\ref{sec:fbperis}, we show that most dependence of the mass loss on orbital parameters can be summarized through its dependence on the pericentre radius of the orbit. As a further simplification, we show in Section~\ref{sec:tidal} that baryonic and DM-only cases follow the same relation when the pericentre tidal field is considered as the primary parameter instead and further, that also the concentration dependence can be explained by defining a single effective tidal field parameter that takes into account the structure-tide degeneracy \citep{2022arXiv220700604S}.

\subsection{On the pericentres }\label{sec:fbperis}

While it is necessary to know the exact orbital configuration of a subhalo and the exact potential structure of the host to make an exact prediction of its mass loss, good approximate predictions can still be obtained only through knowledge of a small sub-set of the parameters. Here, we try to understand what the single most predictive parameter for estimating the mass loss is. First, we investigate the orbital pericentre as a candidate which has been proposed by several other studies~\citep{Penarrubia2010,Drakos2020}).

\begin{figure}
\begin{center}
\includegraphics[width=.48\textwidth]{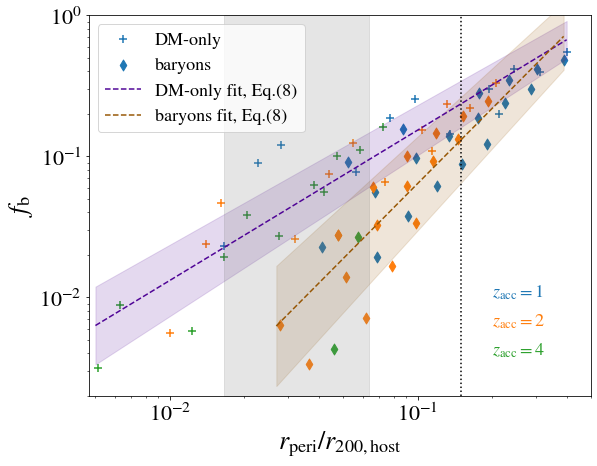}
\includegraphics[width=.48\textwidth]{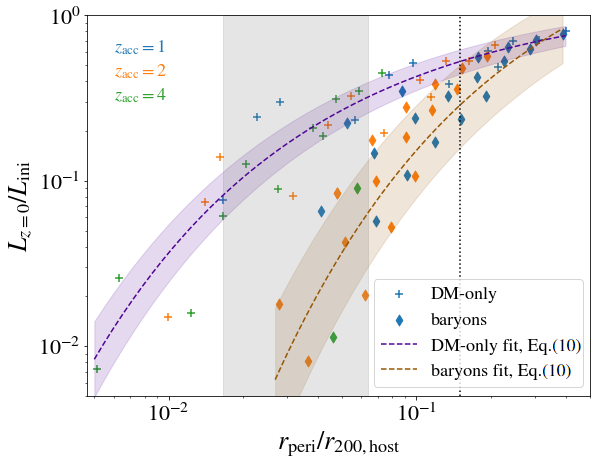}
\caption{
Top: Bound mass fraction at present time as a function of the subhalo pericentric distance in units of the host virial radius at $z=0$, $r_\mathrm{peri}/r_\mathrm{200,host}$, as found in different runs with an initial subhalo concentration $c=10$. Each point corresponds to different orbital parameters and accretion redshifts. Dashed purple and brown lines correspond to fits to Eq.~(\ref{eq:peris}) using the best-fit parameters collected in Table~\ref{tab:peris} for both the cases without and with baryons, respectively. 
Bottom: Annihilation luminosity at present time, normalized to the initial one, as a function of the subhalo pericentric distance in units of the host virial radius, r$_\mathrm{peri}/r_\mathrm{200,host}$, as found in different runs with an initial subhalo concentration $c=10$ and varying the orbital parameters and accretion redshifts. Dashed purple and brown lines correspond to fits to Eq.~(\ref{eq:perisL}) using the best-fit parameters collected in Table~\ref{tab:peris} for both the cases without and with baryons, respectively. 
In both panels, the grey area corresponds to the solar vicinity, defined as the galactocentric region within $8.5 \pm 5$ kpc. The black dotted line shows the radius at which the baryonic tidal field is comparable to the DM halo one; see discussion in Section~\ref{sec:fbperis} and Fig.~\ref{fig:tidalcomps}.
} 
\label{fig:fbperi}
\end{center}
\end{figure}

In the top panel of Fig.~\ref{fig:fbperi} we show $f_\mathrm{b}$ at present time as a function of the pericentre of the orbit\footnote{To be precise, among all pericentres since accretion, we select the one with the minimum distance to the host halo centre. Some small variations are indeed observed among pericentres in the same run, of the order of 10-20\%.} for different orbital parameters and accretion redshifts, We adopt $c = 10$ in all cases.
Our results show that these points are roughly aligned in log-log space: 

\begin{equation}\label{eq:peris}
    f_\mathrm{b} = c  (r_\mathrm{peri}/r_\mathrm{200,host})^m,
\end{equation}
where $r_\mathrm{peri}/r_\mathrm{200,host}$ is the value of the pericentre in each case, i.e. the minimum distance between the subhalo and the host in each simulation, in terms of the virial radius of the host at $z=0$.
Our best fit parameters for those data, for both the non-baryonic and baryonic cases,
are listed in Table~\ref{tab:peris}. The corresponding fits are also shown in the top panel of Fig.~\ref{fig:fbperi} together with their respective scatter.

\begin{table}
	\centering
	\caption{Best-fit parameters and uncertainties for the power law function relating the pericentres with $f_\mathrm{b}$, given by Eq.~(\ref{eq:peris}), and for the function relating the pericentres and $L_\mathrm{z=0} / L_\mathrm{ini}$, described in equation~(\ref{eq:perisL}), both for the cases without and with baryons. These fits are for a particular value of concentrations, namely $c=10$, and are shown in Fig.~\ref{fig:fbperi} together with the data used to perform the fits. See Section \ref{sec:fbperis} for details. 
	}
	\label{tab:peris}
	\begin{tabular}{ccc} %
		\hline
		 & without baryons & with baryons \\
		\hline 
		$m$ & $1.07 \pm 0.07$ & $1.77 \pm 0.16$ \\
		$\log_{10} c$ & $0.25 \pm 0.10$ & $0.58 \pm 0.17$ \\
		\hline 
		$n$ & $1.43 \pm 0.04$ & $3.0 \pm 0.4$ \\
		$d$ & $1.3 \pm 0.1$ & $5 \pm 1$ \\
		\hline
	\end{tabular}
\end{table}

As expected, a smaller pericentre induces a larger mass loss in general. This effect is much greater when baryons are taken into account, since they strongly enhance the tidal field in the centre of the host. 
Interestingly, the scatter is significant in both cases, ranging between $0.13-0.29$ and  $0.54-0.85$ dex for the DM-only and baryonic cases, respectively. This suggests that, even if the pericentric distance is the driving effect in the mass loss, there are other, second order effects also present in the process. It is also worth mentioning that we do not have points for small pericentric distances for the case of including baryons because of the lack of resolution, i.e. these points would lie below our convergence criteria.

\begin{figure}
\begin{center}
\includegraphics[width=\columnwidth]{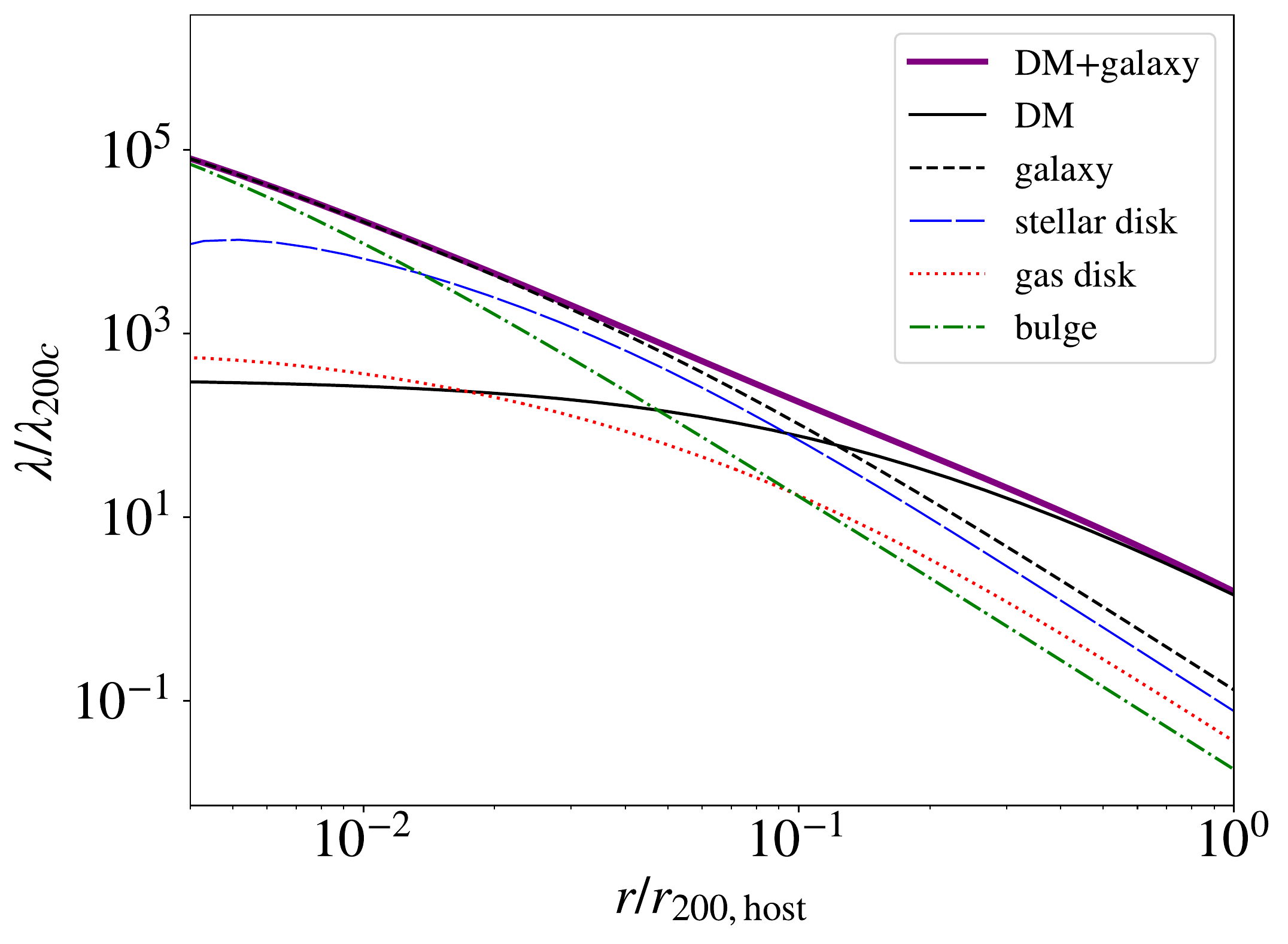}
\caption{Values of the largest eigenvalue of the pericentric tidal field at $z=0$ for the different components with respect to the distance to the host centre, normalized to the host virial radius. We are measuring tidal fields in units of $\lambda_{200}$, i.e. the tidal field that is necessary to create a saddle point in an NFW potential at $r_{200}$ at redshift $z = 0$; see Equation~\ref{eq:lsaddle} for details.
}
\label{fig:tidalcomps}
\end{center}
\end{figure}

Notice that both the non-baryonic and baryonic cases agree when the pericentre is sufficiently large. To get a better understanding of this behaviour, we need to introduce the notion of tidal field. We consider the tidal tensor, $\mathbf{T}$, which has three eigenvalues. The largest of them, $\lambda$, is the most relevant one for our purposes, while the other two might just introduce second order corrections. Fig~\ref{fig:tidalcomps} shows the tidal field in units of\footnote{$\lambda_{200\mathrm{c}}$ is the tidal field that is necessary to introduce a saddle point in an NFW potential at $r_{200}$ at redshift $z = 0$.}
\begin{equation}\label{eq:lsaddle}
    \lambda_{200\mathrm{c}}  = \frac{\partial_r \phi_{{\rm{NFW}}}(r_{200})}{r_{200}} = 100 H_0^2,
\end{equation}
where $H_0$ is the Hubble parameter at redshift $z=0$, as a function of the normalized distance with respect to the host centre. We can see that the tidal field due to baryons is not relevant anymore for $r/r_\mathrm{200,host} \gtrsim 0.2$, since as said they are mostly located in the centre of the host. This explains that results in Fig.~\ref{fig:fbperi} for both the DM-only and baryonic cases are similar for large pericentric distances.

We have done the same analysis for the annihilation luminosity. The bottom panel of Fig.~\ref{fig:fbperi} shows its value at present time normalized to the initial one versus the pericentre of the orbit. We used the same runs that were used for the top panel of the same figure. From this exercise we can estimate the luminosity loss of subhaloes in the solar vicinity, depicted as a grey shaded region in the bottom panel of Fig.~\ref{fig:fbperi}. In particular, if we only consider DM inside the host, subhaloes lose between 70 and 90 per cent of their initial $L$. When we add baryons, this percentage can increase up to 99\%.

We did not find a power-law behaviour in this case. We propose the following fitting function:
\begin{equation}\label{eq:perisL}
    L_\mathrm{z=0} / L_\mathrm{ini} = d \cdot n^{-1/\sqrt{x}},
\end{equation}

Our best-fit parameters are also listed in Table~\ref{tab:peris}. In this case, we observe again that both DM-only and baryonic results converge for large pericentric distances.

While it is intriguing to see that mass loss and luminosity follow simple relations as a function of the pericentre radius, we want to emphasize here that the obtained relations will additionally depend on the initial concentration of the subhalo and on parameters that modify the host potential.

\subsection{Mass loss and the pericentre tidal field}\label{sec:tidal}

As we have seen in Fig.~\ref{fig:fbperi}, the  pericentre versus mass loss relation is different for host potentials that consider baryons and those which do not. This makes sense since tidal fields are much stronger in the host centre in the baryonic cases than in the DM-only case.

In~\citet{2022arXiv220700604S}, we have proposed that both of these cases may be unified into a single relation if we consider their pericentre tidal fields instead of their radii as the important parameters. Additionally, we have proposed in~\citet{2022arXiv220700604S} that the concentration dependence of the tidal stripping problem should additionally disappear if we measure tidal fields in units of the scale tide $\lambda_\mathrm{s}$ and masses in units of the scale mass $M_\mathrm{s}$:
\begin{equation}
\label{eq:Ms}
    M_\mathrm{s} = \frac{\ln(2) - 1/2}{\ln(1+c) - c/(1+c)} M_{200}, \end{equation} \begin{equation} \label{eq:ls}
    \lambda_\mathrm{s} = \frac{\ln(2) - 1/2}{\ln(1+c) - c/(1+c)} c^3 \lambda_{200}.
\end{equation}

In~\citet{2022arXiv220700604S}, we have developed a simple model that describes NFW haloes that are exposed to a tidal field, the latter increasing so slowly that the halo responds adiabatically. In the adiabatic limit (and assuming an isotropic tidal field), $M_\mathrm{b} / M_\mathrm{s}$, where  $M_\mathrm{b}$ is the remaining mass in such limit, is exactly only a function of the effective tidal field, $\lambda / \lambda_\mathrm{s}$. Now, in realistic setups many additional dependencies exist, but we would still expect that at first order most of the host potential dependence and most of the concentration dependence should disappear if results are presented in this way. Here we want to test this expectation.

We measure the three eigenvalues $\lambda_1 \geq \lambda_2 \geq \lambda_3$ of the tidal tensor that the subhalo is exposed to at each timestep. Then, we infer the maximum value of $\lambda_1$ among all of the timesteps and we define this value as the pericentre tidal field, $\lambda_{\rm{peri}}$.\footnote{We note that, in the baryonic case and because of the galactic disk, this maximum tidal field may be reached at a point that does not exactly correspond to the actual pericentre, yet it will be typically very close.} Using the maximum of the tidal field as $\lambda_\mathrm{peri}$ has the advantage that it is always well defined even in cases of anisotropic or evolving host potentials, etc. 

\begin{figure}
\begin{center}
\includegraphics[width=0.96\columnwidth]{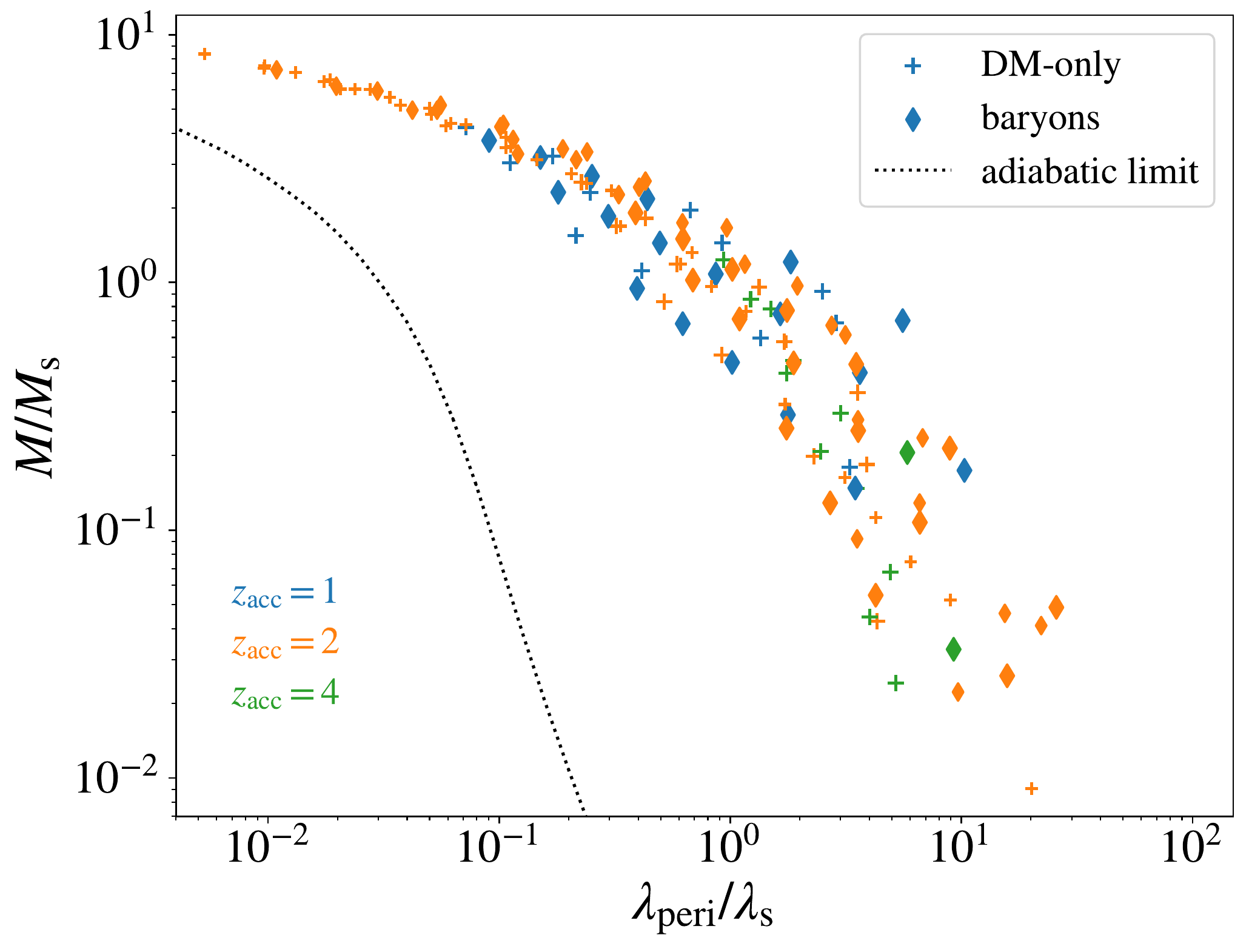}
\includegraphics[width=\columnwidth]{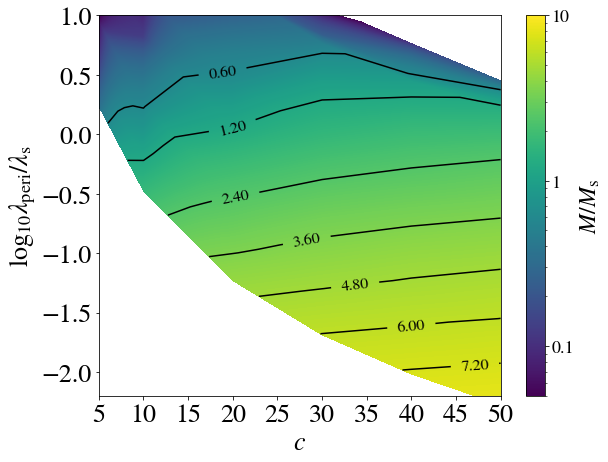}
\caption{Top:
Subhalo mass at present divided by the initial scale mass (Eq.~\ref{eq:Ms}), as a function of the effective tide, i.e. the largest tidal tensor eigenvalue at the pericentre divided by the scale-tide $\lambda_\mathrm{s}$ (Eq.~\ref{eq:ls}), for different subhalo concentrations (for $z_\mathrm{acc} = 2)$, redshifts (for $c=10$) and orbital parameters, both with (diamonds) and without (plus signs) baryons. The dotted line is the adiabatic limit of \citet{2022arXiv220700604S}, which represents the absolute maximum mass loss expected after an infinite number of orbits.
Bottom: Remaining subhalo mass at present time divided by the scale mass, for different concentrations and effective tides at the pericentre. Iso-contours are now almost horizontal in contrast to those in Fig.~\ref{fig:fbsurf}. Note that both the bottom left and top right corners are in blank because they cannot be populated with our simulation suite (described in Table~\ref{tab:fiduset}).
} 
\label{fig:surfbarytid}
\end{center}
\end{figure}

We show the mass loss $M/M_\mathrm{s}$ as a function of $\lambda_\mathrm{peri} / \lambda_\mathrm{s}$ in the top panel of Fig.~\ref{fig:surfbarytid}, where we have combined runs with different concentrations, orbital parameters and accretion redshifts, both with and without baryons.  Strikingly, the cases with and without baryons follow the same relation when shown in this manner. This shows that the largest encountered tidal field is indeed the single most important parameter for understanding tidal mass loss. Of course, there is a sizeable scatter in the relation, which shows that secondary dependencies exist.\footnote{In particular, we noticed that i) $x_\mathrm{c}$ is the driving parameter producing the vertical scatter, in such a way that for the same pericentric distance, different $x_\mathrm{c}$ values give a significantly different mass loss even for the same values of $\eta$; 
ii) the horizontal scatter is explained since subhaloes in more circular orbits and with a small $x_\mathrm{c}$ are closer to the host centre for longer times compared to subhaloes in radial orbits with large $x_\mathrm{c}$ values: the former ones have larger pericentres but suffer the same mass loss in the end.} Yet, the relation is now considerably tighter than the one shown in the top panel of Fig.~\ref{fig:fbperi}. In the same top panel of Fig.~\ref{fig:surfbarytid}, we show a line corresponding to the adiabatic limit of ~\citet{2022arXiv220700604S}, which represents the absolute maximum expected mass loss in this parameter space. We note that our measured values here still lie quite far from the adiabatic limit. This is expected, as these subhaloes have orbited for much shorter times than what is necessary to reach the mentioned limit. Additionally, we may be overestimating the pericentre tidal field here a bit, by taking the maximum across the full history.

In the bottom panel of Fig.~\ref{fig:surfbarytid} we show the mass loss as a function of the effective tide and concentration. When presented in these reduced units, the concentration dependence indeed disappears, i.e. the iso-contours in this plot are approximately horizontal. This shows that much of the parameter space of the tidal mass loss problem can be simplified and generalized. The dependence on the initial concentration is degenerate with the dependence on the amplitude of the tidal field and we can summarize these two into one effective parameter $\lambda_\mathrm{peri} / \lambda_\mathrm{s}$ -- we call this phenomenon the ``structure-tide'' degeneracy and we explain in~\citet{2022arXiv220700604S} how this arises from the invariance of the Vlasov-Poisson system to time-rescalings. Note that we are not able to populate the lower left corner of this surface plot (bottom panel of Fig.~\ref{fig:surfbarytid}) since we cannot reach such low values of the effective tide for small concentrations with the orbital configurations we are allowing in this work. A similar situation occurs for large tidal fields and concentrations.

The adiabatic limit prediction of this plot can be seen in the top left panel of Fig. 14 in \citet{2022arXiv220700604S}. Again, we note that the here considered scenarios have lost much less mass than the adiabatic limit in \citet{2022arXiv220700604S}, as expected (see explanation above).

\section{Conclusions}\label{sec:conclu}

Cosmological \nbody simulations are computationally expensive and they are prone to both mass and spatial resolution limits, which makes it very difficult to properly resolve subhaloes and follow their evolution within their hosts. In contrast, employing an analytical prescription when modelling the host halo potential gives plenty of room to simulate a subhalo and to track its evolution with great accuracy and numerical resolution.

This work makes use of DASH, a code specifically designed to perform this task with unprecedented accuracy, reaching solar-mass and sub-parsec resolution in our simulations. In particular, in this work we have implemented a few, important novelties with respect to the original version in~\citet{Ogiya2019}, which have made our results more realistic and useful. The most relevant ones are the inclusion of the evolution of the DM host potential; the implementation of the baryonic potential, which also evolves with time; and a new routine to select those orbital parameters that lead to a greater probability for a subhalo to cross the solar galactocentric radius, i.e. those potentially most relevant for DM searches.

We have studied the evolution of subhaloes in a MW potential, the latter described as an NFW DM halo plus three baryonic components (stellar and gas disks, and bulge). We have explored different subhalo configurations, adopting a fiducial set of parameters as the representative case, but also varying one or some of these parameters to understand the role of each of them in the evolution of the subhalo. We have focused on studying two quantities particularly relevant for our purposes, the bound mass fraction and the DM annihilation luminosity. We have also performed several important convergence checks that allow us to confidently derive robust conclusions. Our main findings can be summarized as follows:

\begin{itemize}

\item[$\star$] Contrary to~\citet{Kelley2019, 2021MNRAS.501.3558G}, we find that subhaloes do survive in the innermost 15~kpc of our galaxy, although they typically lose more than 90\% of their initial masses (see top panel of Fig.~\ref{fig:fbperi}).

\item[$\star$] Subhaloes with lower concentrations and subhaloes on orbits with smaller pericentric distances are more depleted. Similarly, subhaloes accreted earlier or with lower orbital energies have smaller orbits and have lost more mass at $z=0$. This is illustrated, e.g., in Fig.~\ref{fig:fb1}. Including baryonic material in the host induces a significantly larger mass loss in most cases as well, e.g. an order of magnitude more in the fiducial case (see e.g. left bottom panel of Fig.~\ref{fig:fbsurf}).

\item[$\star$] Subhaloes in parallel orbits with respect to the galactic disk lose significantly more material than those orbiting in more perpendicular orbits, similarly to that found in~\citet{2021MNRAS.tmp.2848G}. Yet, the latter still lose significantly more mass than subhaloes orbiting a DM-only host. We also report more substantial mass loss for parallel orbits compared to~\citet{2021MNRAS.tmp.2848G}; see the right panel of Fig.~\ref{fig:fullbary}. Indeed, our results suggest that the relevance of disk shocking may be negligible compared to the undergoing baryonic tidal field during the whole evolution of the subhalo. 

\item[$\star$] Subhaloes orbiting a DM-only halo with a pericentre in the solar vicinity have lost 70-90\% of their initial annihilation luminosity at $z=0$. This percentage increases up to 99\% when baryons are included in the host (bottom panel of Fig.~\ref{fig:fbperi}). In other words, we expect nearby low-mass subhaloes to be around ten times less luminous with respect to those in DM-only.

\item[$\star$]  
We emphasize that our results are virtually independent of subhalo mass for subhaloes lighter than $10^8\, \msun$ (Fig.~\ref{fig:fullbary}). This was already stated by \citet{Ogiya2019} for the DM-only case and it is now confirmed for the baryonic scenario as well.

\item[$\star$]  We have found new ways of summarizing the most important dependencies on the parameter space into a single parameter. Firstly, we have found that the orbital dependence of subhalo mass loss can be summarized at first order into its dependence on the pericentre radius. We have found simple powerlaw relations for a $c=10$ subhalo that orbits in a Milky-Way like host --with different relations for baryonic and DM-only cases. We note that these relations are not general though, but rather hold only for the specific concentration $c=10$ that we investigated.

\item[$\star$] Motivated by the analytical arguments of \citet{2022arXiv220700604S}, we have additionally found that the problem can be further simplified, by summarizing the concentration and host-potential dependence  into the single parameter $\lambda / \lambda_s$ --the effective tidal field at pericentre. The host-potential dependence (e.g. baryons versus DM-only) is captured, by using the pericentre tidal field instead of the pericentre radius, since the tidal field is ultimately the cause of the mass loss. Further, the concentration dependence is captured, by normalizing to the scale tide $\lambda_s$ which depends on the concentration and is degenerate in its effects with the amplitude of the tidal field $\lambda$. We refer to this as the ''structure-tide'' degeneracy and explain in \citet{2022arXiv220700604S} how it naturally arises from the time-rescaling invariance of the Vlasov-Poisson system.

\end{itemize}

Studying subhalo survival is crucial to elucidate the role of small subhaloes in indirect DM searches, which was one of the key motivations to perform this work.
Among potential future applications of our work we can mention, for instance, a more refined calculation of the so-called subhalo boost factor to annihilation signals, more robust constraints on DM, especially for those scientific cases where subhaloes play a central role, and the optimization of DM search observation strategies for spatially extended DM targets. Some of these applications are already under study and will be presented elsewhere.

This work is still ongoing. In the near future, we will 
take a closer look at the evolution of the subhalo concentration with time, as well as the impact of the latter for indirect DM searches.  
We would also like to understand the impact of our findings on both the radial distribution and mass function of the MW subhalo population. Besides, we are considering running more massive simulations with higher resolution, which will allow to track $f_\mathrm{b}$ for more extreme $c$ cases, as well as calculating the annihilation luminosity with higher confidence.

\section*{Acknowledgements}


The work of AAS and MASC was supported by the grants PGC2018-095161-B-I00 and CEX2020-001007-S, both funded by MCIN/AEI/10.13039/501100011033 and by ``ERDF A way of making Europe''. The work of AAS was also supported by the Spanish Ministry of Science and Innovation through the grant FPI-UAM 2018. MASC was also supported by the Atracci\'on de Talento contract no. 2020-5A/TIC-19725 granted by the Comunidad de Madrid in Spain. 
GO was supported by the Fundamental Research Fund for Chinese Central Universities (Grant No. NZ2020021) and the Waterloo Centre for Astrophysics Fellowship. 
JS and RA acknowledge the support of the European Research Council through grant number ERC-StG/716151 (``BACCO'').

Our simulations were carried out in the Atlas and Graham supercomputers, operated by the DIPC and Compute Canada (\url{www.computecanada.ca}), respectively.

This research made use of Python, along with community-developed or maintained software packages, including IPython \citep{Ipython_paper}, Matplotlib \citep{Matplotlib_paper}, NumPy \citep{Numpy_paper} and SciPy \citep{2020SciPy-NMeth}.




\bibliographystyle{mnras}

\bibliography{References.bib}



\appendix

\section{A deeper look into convergence}\label{sec:apconv}


\subsection{Bound mass fraction}\label{sec:apfb}

We have taken into account two different numerical convergence criteria in order to elucidate whether the subhalo has been physically or numerically disrupted. The first one depends on the softening length, $\varepsilon$, and the second is related to the number of particles, $N$~\citep{vandenBosch2018_num_criteria}. The maximum among the two of these criteria for a given run is the one we finally adopt in each case. In particular:

\begin{multline}\label{eq:boundcrit}
f_\mathrm{b} > f_\mathrm{b}^\mathrm{min,1} = 1.12 \frac{c^{1.26}}{f^2(c)} \left(\frac{\varepsilon}{r_\mathrm{s,0}} \right)^2 \\
f_\mathrm{b} > f_\mathrm{b}^\mathrm{min,2} = 0.32 \left(\frac{N_\mathrm{acc}}{1000} \right)^{-0.8} \\
f_\mathrm{b}^\mathrm{min} \equiv \max\left(f_\mathrm{b}^\mathrm{min,1}, f_\mathrm{b}^\mathrm{min,2}\right)
\end{multline}

We consider that the subhalo has been numerically disrupted and no robust conclusions can be obtained from that simulation if the bound mass fraction drops below that value before $z=0$. We note though that this line lies always below $10^{-2}$, i.e. the subhalo has already lost more than 99\% of its mass by then. Nonetheless, this does not necessarily mean that the subhalo has been \textit{physically} disrupted, but rather we do not have enough resolution to study accurately the mass loss beyond.

\begin{figure*}
\begin{center}
\includegraphics[width=.495\textwidth]{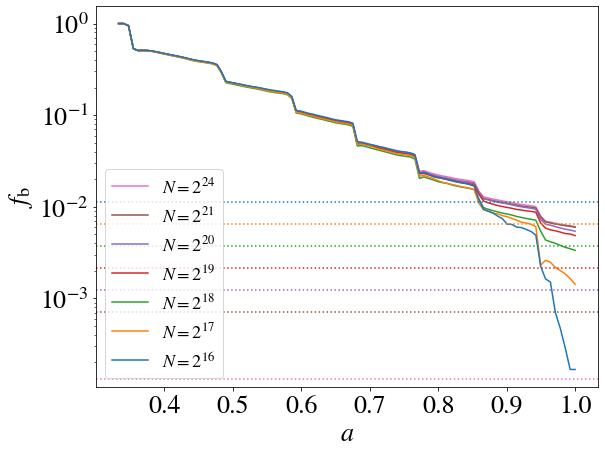}
\includegraphics[width=.495\textwidth]{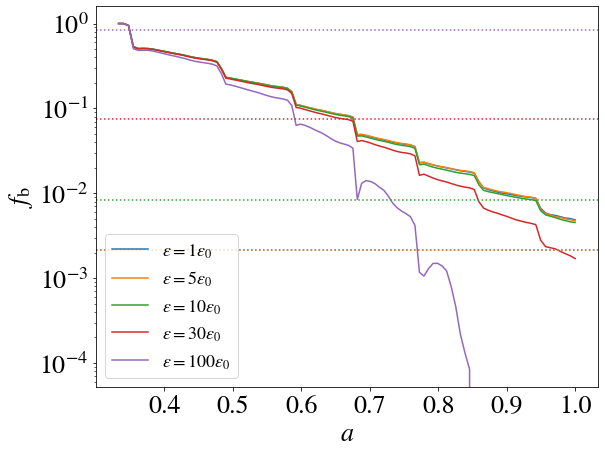}
\caption{Convergence checks including baryons. 
Left panel:
 Bound mass fraction for different initial number of particles, $N$, together with their corresponding convergence limits, the latter depicted as horizontal dotted lines coloured according to the legend. 
Right panel:  Bound mass fraction for different values of the softening length $\varepsilon$, where $\varepsilon_0=0.0003\times r\sub{200,sub} = 0.304$ pc in the legend.
We adopt our fiducial setup of Table~\ref{tab:fiduset} in all cases and the inclination angle is fixed to 45 deg in the baryonic runs.
}
\label{fig:apfullbary}
\end{center}
\end{figure*}

These criteria in equation~\ref{eq:boundcrit} have been tested for runs without baryons in~\citet{vandenBosch2018_num_criteria}. Here, we have verified that they still hold when the baryonic components are added to the host potential as well, by performing several simulations employing our default setting and changing the numerical parameters, $N$ and $\varepsilon$.
The results are shown in~Fig.~\ref{fig:apfullbary}, 
where we can see that the mass loss is increased when either criterion is unsatisfied
and that the criteria work nicely with baryons, too, since the results converge --within a small scatter-- when they are above their respective convergence line. We have also found that increasing $\varepsilon$ can result in less mass loss even below the convergence line. This can be explained since the two-body relaxation timescale also depends slightly on $\varepsilon$; the smaller $\varepsilon$, the shorter relaxation timescale. We will discuss the effect of relaxation timescale below. 

\begin{figure}
\begin{center}
\includegraphics[width=.43\textwidth]{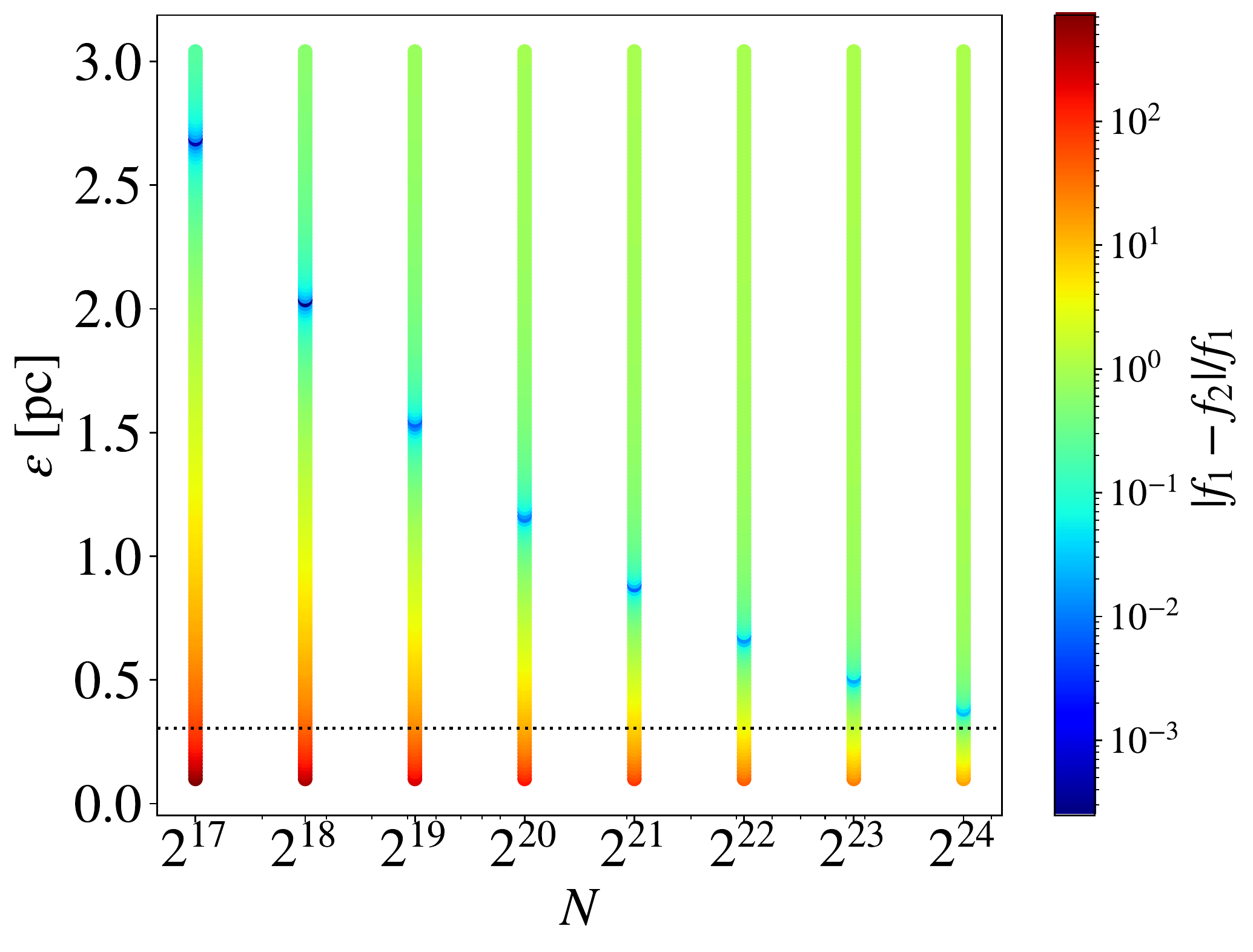}
\includegraphics[width=.43\textwidth]{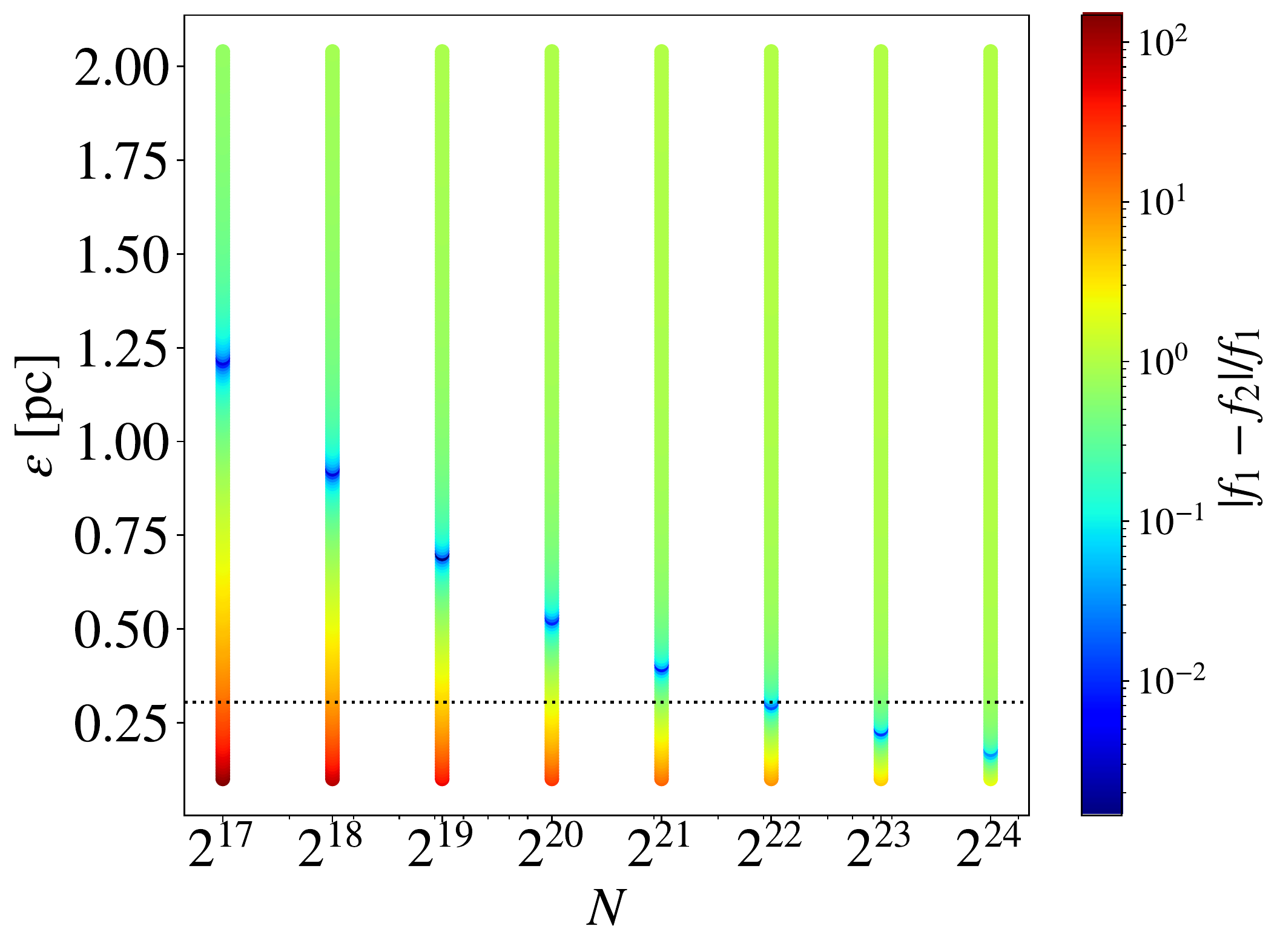}
\caption{Ratio between both $f_\mathrm{b}$ convergence criteria in Eq.~(\ref{eq:boundcrit}), for different pairs $N-\varepsilon$ and for two initial subhalo concentrations, $c=10$ (top) and $c=20$ (bottom). The dotted line corresponds to our default $\varepsilon$ for $z_\mathrm{acc} = 2$. See text for details. 
} 
\label{fig:fboundcrit}
\end{center}
\end{figure}

It is important to realize that a larger $N$ or a smaller $\varepsilon$ costs more computational time. 
Choosing the optimal $N$ and $\varepsilon$ values for each simulation is a non trivial but an important task, since we need to find a compromise between computational time and numerical resolution. Since the critical value of $f_\mathrm{b}$ is the maximum of the two criteria in equation~\ref{eq:boundcrit}, we can compute when these values are closer depending on our numerical parameters. We have plotted this relation between the criteria in Fig.~\ref{fig:fboundcrit} for two different concentration values, $c=10$ and $c=20$. We are only showing $N$ values which are a power of 2 because we are using those in this work. The colourbar represents how close these values are, where the dark blue means they are the closest, i.e. a smaller difference between both criteria, which is optimal for our purposes. However, increasing $N$ makes the optimal $\varepsilon$ smaller. Therefore, the convergence value is not going to improve below the dark blue points. The same happens to the right.

In our work, we use a different number of particles depending on the case we are studying. On one hand, our $N$ ranges from $2^{18}$ to $2^{21}$. We have also studied the fiducial case with larger values of $N$, up to $2^{24}$, to verify the convergence of our results. On the other hand, we are going to set $\varepsilon=0.0003\,r\sub{200,sub}$ 
from now on, 
which is good enough for our purposes.

\begin{figure}
\begin{center}
\includegraphics[width=\columnwidth]{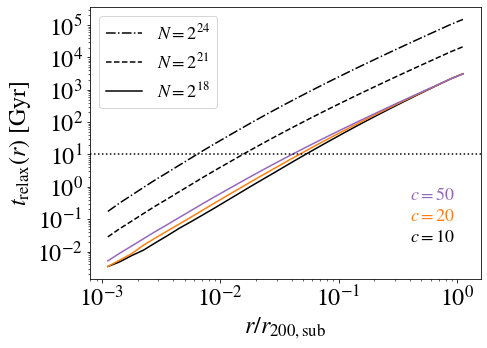}
\caption{Relaxation timescale for different concentrations and number of particles, using the initial snapshot for runs with $z_\mathrm{acc} = 2$, as a function of the normalized subhalo inner radius. The horizontal dotted black line corresponds to the time of the simulation, $t_\mathrm{sim}$, 
and indicates the radius above which the simulation can be trusted. 
}
\label{fig:trelax}
\end{center}
\end{figure}

As stated above,
we also have to take into account the relaxation timescale. When a system is described with a finite number of particles, the acceleration of each one eventually deviates from the mean value when particles get close to each other~\citep{2003MNRAS.338...14P, 2008gady.book.....B}. These `collisions’ drive changes of order unity in energy on the relaxation timescale, which is obtained as:

\begin{equation}
    t_\mathrm{relax} \simeq \frac{N(<r)}{8 \ln N(<r)} t_\mathrm{cross}(r),
\end{equation}

where $t_\mathrm{cross}(r) = \frac{2 \pi r}{V_\mathrm{c}(r)} = 2 \pi \sqrt{\frac{r^3}{GM(r)}}$ is the crossing time, a rough estimation of the orbital period of a particle at $r$ in the subhalo.
After one relaxation time, the cumulative small kicks from many encounters with other particles have changed the particle's 
orbit significantly from the one it would have had if the gravitational field had been smooth, meaning the particle 
has lost its memory of its initial conditions when a relaxation time has passed. This implies that the larger the relaxation time, the more trustful the results obtained at the innermost parts. More specifically, we can trust a simulation at radius $r$ if it satisfies $t_\mathrm{relax} (r) > t_\mathrm{sim}$, where $t_\mathrm{sim}$ depends on $z_\mathrm{acc}$. 
This relaxation timescale is shown in Fig.~\ref{fig:trelax} for different radii, subhalo concentrations and number of particles, using the first snapshot in every case. It is slightly larger for a larger concentration or number of particles, which means we may need to use more particles when the concentration is smaller. 
More specifically, we can trust our data from $r / r_\mathrm{200,sub} \sim 5 \cdot 10^{-3}$ if $N = 2^{24}$ and $z_\mathrm{acc} = 2$ while, if $N = 2^{18}$, the simulations are believable only from $r / r_\mathrm{200,sub} \sim 3 \cdot 10^{-2}$. This will be particularly relevant when the subhalo experiences a significant mass loss.

\subsection{Annihilation luminosity}\label{sec:aplu}

\begin{figure}
\begin{center}
\includegraphics[width=\columnwidth]{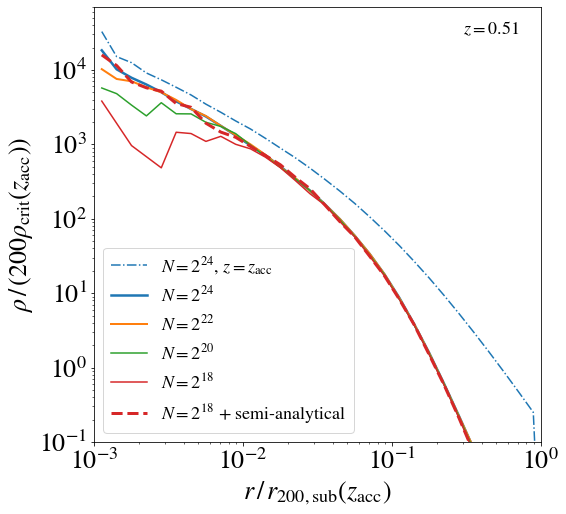} 
\caption{
Example of subhalo DM density profile $\rho$, normalized to 200 times the critical density at accretion, after several pericentric passages for runs with a different total number of particles, $N$.  The red dashed line corresponds to the hybrid, semi-analytical scheme we adopted to recover the cusp in low-resolution runs in order to save computational time; see Appendix~\ref{sec:aplu} for details.
} 
\label{fig:aplumis}
\end{center}
\end{figure}

As explained in Section~\ref{sec:lumi}, we are facing numerical resolution issues when studying the evolution of the innermost regions of the subhalo. Indeed, when  $N$ is not large enough, the subhalo inner cusp ends transforming into a core without any physical explanation. This can be seen in 
Fig.~\ref{fig:aplumis}, where we show an example of a density profile for different values of $N$ and our fiducial set of parameters, including baryons, some time after the subhalo has been accreted. For a considerably large number of particles, $N = 2^{24}$ in our case of study, we mostly recover a cusp in the subhalo centre, while this is not the case for a lower number of particles. Note that the subhalo gets truncated in its outskirts as it orbits around the host and loses mass, as physically expected, and this behaviour does not change for a lower resolution run. 

Unfortunately, working with such large resolutions becomes unsuitable in terms of the computational cost. Because of this, we have implemented a hybrid approach to recover the inner cusp even in cases where the resolution is not sufficiently high. To do so, we first set a critical subhalo radius, $x_\mathrm{crit} = r_\mathrm{crit} / r_\mathrm{200,sub}$, 
 the radius from which we can trust our results according to the relaxation timescale for a given number of particles (see Appendix~\ref{sec:apfb}), and trust the simulation data only beyond that point ($x > x_\mathrm{crit}$). As for the innermost subhalo region, which we recall is of special relevance for our DM annihilation studies, we use the semi-analytical model in \citet{2019MNRAS.490.2091G}  to describe the evolution of the subhalo inner cusp with time, down to $x = 5 \cdot 10^{-3}$, which corresponds to the relaxation timescale for the higher resolution we have tested, $N = 2^{24}$ particles. The red dashed line in Fig.~\ref{fig:aplumis} shows how this hybrid scheme fixes the internal part of the subhalo, giving a similar result as the high-resolution, $2^{24}$ particles run.

\bsp	
\label{lastpage}
\end{document}